\documentclass[11pt,twoside
]{article}
\pdfoutput=1
\RequirePackage[left=30mm,right=25mm,textheight=235mm,top=25mm,footskip=10mm]{geometry}
\RequirePackage{fancyhdr}  
\RequirePackage{empheq}

\usepackage{sty/artcljr}

\newcommand{\coldb}[1]{\bgroup\color{red}{#1}\egroup}
\newcommand{\coljr}[1]{\bgroup\color{blue}{#1}\egroup}

\begin{document}
 
\thispagestyle{empty}


\begin{figure}[htb]
\unitlength1cm
\begin{picture}(21,1.25)
%
\put(0.0,-1.0){
\put(-2.0, 3.1){{\sffamily\textbf{\Large Ruhr-Universit\"at Bochum}}}
\put(-2.0, 2.5){{\sffamily\textbf{Chair of Continuum Mechanics}}}
\put(-2.0, 1.90){{\sffamily{Prof.~Dr.-Ing.~habil. Daniel Balzani}}}
}
\end{picture}
\end{figure}


\vspace{3cm}

{\Huge
A Simple and Efficient Lagrange Multiplier Based Mixed Finite Element for Gradient Damage\\
} 

\vspace{5mm}

J. Riesselmann and D. Balzani

\vfill 

\title{A Simple and Efficient Lagrange Multiplier Based Mixed Finite Element for Gradient Damage}

\author{J. Riesselmann $^1$ and
        D. Balzani $^{*,1}$}

\address{
$^1$ Chair of Continuum Mechanics,
     Ruhr University Bochum,    
     Germany
}

\corresp{
daniel.balzani@rub.de
\vspace{-0.2cm}
}

\abstract{
A novel finite element formulation for gradient-regularized damage models is presented which allows for the robust, efficient, and mesh-independent simulation of damage phenomena in engineering and biological materials. 
The paper presents a Lagrange multiplier based mixed finite element formulation for finite strains. 
Thereby, no numerical stabilization or penalty parameters are required. 
On the other hand, no additional degrees of freedom appear for the Lagrange multiplier which is achieved through a suitable FE-interpolation scheme allowing for static condensation. 
In contrast to competitive approaches from the literature with similar efficiency, the proposed formulation does not require cross-element information and thus, a straightforward implementation using standard element routine interfaces is enabled. 
Numerical tests show mesh-independent solutions, robustness of the solution procedure for states of severe damage and under cyclic loading conditions. 
It is demonstrated that the computing time of the gradient damage calculations exceeds the one of purely elastic computations only by an insignificant amount. 
Furthermore, an improved convergence behavior compared to alternative approaches is shown. 
}


\acknowledge{
The authors greatly appreciate financial funding by the German Science Foundation (Deutsche Forschungsgemeinschaft, DFG), as part of the Priority Program 
2256 ("Variational Methods for Predicting Complex Phenomena in Engineering Structures and Materials"), project ID 441154176, reference ID BA2823/17-1.
}

\maketitle

 \vspace{-0.8cm}
 \section{Introduction \label{sec:introd}}
\vspace{-0.2cm}
The reduction of stiffness and softening of engineering- and biological materials due to deterioration of the microstructure can be described via continuum damage models.
In these models, a damage variable is introduced whose evolution is governed through corresponding constitutive relations.
The evolution of the damage variable contributes to a lowering of the strain energy density accounting for the material softening behavior (see \cite{Lem:1984:htu,LemCha:1990:mos} and \cite{Kac:1986:itc} for small strains).
Finite strain damage formulations can be found in the works \cite{Mie:dac:1995} and \cite{MenSte:2001:ata} (see also \cite{BalSchGro:sod:2006,BalBriHol:2012:cff} for application to the damage modeling of biological tissue).
While for relatively small amounts of damage, corresponding numerical schemes of local damage formulations may render sufficiently accurate solutions, the governing equations may lose ellipticity upon higher damage intensities.
Consequently, in these cases, the numerical solution algorithms fail to converge and mesh-dependent results are obtained due to localization of numerical values of the damage variable.

Thus, various regularization methods exist to avoid this issue and the development of robust and compuationally efficient numerical formulations are a field of ongoing research.
One approach is to construct a relaxed incremental stress potential, which convexifies the original nonconvex problem (cf \cite{GueMie:2011:oed}, see also the large strain approach of \cite{BalOrt:2012:riv,SchBal:2016:riv} and \cite{Mie:2011:amf} for application to multi-field dissipative problems).
Such relaxation schemes have originally not been able to model strain-softening which is why these approaches have not attracted much attention. 
It is only recently that extended formulations have shown to allow for strain softening, see \cite{SchJunHac:2020:vro} and \cite{KoeNeuMelPetBal:2022:aco}.
Another approach is spacial regularization through gradient enhancement of either the damage variable or the damage driving quantities (cf. \cite{PeeBorBreVre:1996:ged}, see also \cite{PlaBarMisTim:2021:mbe}, which introduces a strain gradient damage formulation for materials with granular microstructure).
In the former case,
through the introduction of the gradient enhancement term, the damage evolution equation transitions from a local constant equation to a partial differential equation.
As a consequence of the higher regularity, the loss of ellipticity can be avoided and mesh-independent solutions can be obtained.
However, because of the continuity requirement of the damage variable, a simple update of the damage equation at the material point level is not directly possible anymore.
If the partial differential equation describing the damage evolution is considered in strong form, an update at the material point level remains possible.
The corresponding field equation in this case, however, contains the Laplace operator, for which suitable discretization schemes are not straightforward.
The work of \cite{JunSchJanHac:2019:afa} introduces a corresponding numerical scheme for small strains (in this context see also \cite{VogJun:2019:aah}, which has been extended to finite strains in \cite{JunRieBal:2021:ent}.
The major drawback of these approaches is that the classical material subroutine interface is not sufficient and information across elements is required rendering the implementation difficult. 
An alternative overcoming this drawback is to introduce a mixed finite element formulation with an additional nodal solution variable in order to include the gradient enhancement.
One of the earlier corresponding contributions is the finite element formulation of \cite{PeeBorBreVre:1996:ged}, in which the damage update is driven by an equivalent strain measure, which is introduced as mixed variable, resulting in a two-field mixed formulation.
Another approach can be found in \cite{LieSteBen:2001:tac,DimHac:2008:amf} and the finite strain formulation of \cite{WafPolMenBla:2013:age}.
There, the additional nodal variable corresponds to the damage field itself and equivalence to the locally updated history parameter is enforced via a penalty constraint term.
The penalty term is also referred to als micromorphic coupling term (cf. \cite{For:2009:maf}).
Since corresponding solution schemes can be solved monolithically while the active set search is governed by a local history parameter, no additional global iterations as e.g. in staggered approaches are necessary (cf. phase field formulations \cite{MieWelHof:2010:tcp,MieHofWel:2010:apf,GerDeL:2019:opi} similar to gradient damage).
Yet, for mixed gradient damage finite element formulations, numerical robustness at severe damage remains a challenge.
In this case the stiffness contributions of some corresponding elements may be significantly lowered, leading to challenging global matrix conditions, which become even more challenging when a penalty term is present.

Therefore, in this contribution we introduce a Lagrange mutliplier based mixed finite element formulation, whose interpolation functions are known to fulfill numerical stability conditions while evading the need for a penalty term.
Moreover, due to a discretization setup which enables for static condensation the size of the resulting global tangent matrix is kept at a minimum, zero-valued diagonal submatrices are avoided, and a positive definite symmetric global matrix can be obtained.
The structure of this contribution is as follows.
In section~\ref{s:framework} the continuum mechanical framework for the modeling of gradient damage at finite strains is given.
The following section~\ref{s:fem} then introduces the proposed mixed finite element formulation.
Here, in subsection~\ref{ss:continuous_formulation} the formulation is presented in the continuous setting, followed by the corresponding finite element discretization, matrix formulation and algorithmic treatment in subsection~\ref{ss:discretization}.
Finally, in section~\ref{s:tests} the results of various numerical studies testing the robustness and convergence properties of the proposed formulation are shown.

%
%
%
%
  \section{Continuum mechanical framework\label{s:framework}}
This section covers the continuum mechanical framework, in which the proposed formulation of the following section is embedded.
First, section \ref{ss:fundamentals} introduces some fundamentals with respect to finite strain kinematics and thermodynamics.
Then, in section \ref{ss:cont_damage_mech} the finite strain gradient damage boundary value problem is presented.

\subsection{Fundamentals\label{ss:fundamentals}}
Let $\phi:\B\rightarrow \Bt$ denote the finite deformation map, where $\B$ is the body in reference configuration and $\Bt$ is the body in the deformed configuration.
The deformation gradient is denoted by $\bF\defeq\Grad\phi= \Grad\bu+\bone$, where $\bu$ is the displacement and $\Grad(\bullet)\defeq\p{\bX}(\bullet)$ denotes the gradient with respect to the reference coordinates.
In the context of continuum damage modeling at finite deformations (cf. \cite{Mie:dac:1995,MenSte:2001:ata}, see also \cite{BalBriHol:2012:cff,BalOrt:2012:riv}) the strain energy density function can be written as follows:
\eb
\sed(\bF,\iv)\defeq (1-D(\iv))\sed_0(\bF).
\label{e:1-D_strain_energy}
\ee
Here, $\sed_0$ denotes a fictitiously undamaged, objective (that is $\sed_0\defeq\sed_0(\bF^T\smpc\bF)$, cf. \cite{Wri:2008:nfe}) hyperelastic strain energy density and the function
$D:\bR^+\rightarrow [0,1)$ describes the material softening due to the evolution of the internal damage variable $\iv$.
For the damage function the value $D(0)=0$ corresponds to the completely intact state and with $D\rightarrow 1$ the material approaches the completely damaged state.

We denote $\Pi$ as  the total energy potential defined over $\B$ and $\Gamma$ as the total entropy defined over $\B$.
The first and second law of thermodynamics read
\eb
 \dot{\Pi} = 0 \qquad
 \text{and} \qquad
 \dot{\Gamma}  \geq 0,
 \footnote{$\dot{(\bullet)}\defeq\p{t}{(\bullet)}$ denotes the time derivative.}
 \label{e:thermo_laws_fundamental}
\ee
which need to be fulfilled independent of the rates of the process variables.
In the context of this contribution process conditions are assumed do be adiabatic, isothermal and quasi static.
Thus, the energy potential simplifies to the sum of the internal energy $\Pi_{\mrm{int}}$ and the potential of external work forces $\Pi_{\mrm{ext}}$.
Process variables are the displacements $\bu$ and the damage variable $\iv$.

\subsection{The Gradient Damage Problem\label{ss:cont_damage_mech}}
The gradient extended potential energy and the dissipation potential read:
%
\eb
\begin{array}{rcl}
\Pi &=& \underbrace{\int_{\B} \Big[ \sed(\bF,\iv)+ \frac{\nlpar}{2} \Grad\iv \scp \Grad\iv +\diss(\iv)\Big] \dX}_{\Pi_{\mrm{int}}}
\underbrace{-\int_{\B}  \bu \scp \bbf \dX - \int_{\dBN} \bu \scp \bt \dA }_{\Pi_{\mrm{ext}}},
\\
\Gamma &=& \int_{\B} \diss(\iv) \dX,
\end{array}
\label{e:total_dissipation_potential}
\ee
with the body forces $\bbf\in L^2(\B)$ and the surface tractions $\bt\in L^2(\dBN)$ on the Neumann boundary $\dBN$.
The integrand of the internal potential energy $\Pi_{\mrm{int}}$ consists of the sum of the strain energy density $\sed(\bF,\iv)$, the function $\diss(\iv)$, which models the dissipation due to evolution of the microstructure
and the gradient term $\nlpar/2\, \Grad\iv \scp \Grad\iv$, which serves as regularization to obtain mesh-independent solutions.
Here, $\nlpar>0$ denotes a nonlocal material parameter. 
The function $\diss$ is chosen to be a continuous monotonic increasing function $\diss:\R^+\rightarrow \R^+$ (with $\p{\iv}{\diss}\geq 0$ so that \eqref{e:thermo_laws_fundamental}$_2$ is always fulfilled if $\dot{\iv}\geq0$ (see e.g. \cite{MieWelHof:2010:tcp}).
Insertion of 
\eqref{e:total_dissipation_potential} into \eqref{e:thermo_laws_fundamental} yields the following strong form of the gradient damage problem (cf. appendix \ref{app:strongf_initial_problem}):
Find $\iv$ and $\bu$ such that
\eb
\begin{array}{rcl}
 -\Div\bP&=&\bbf \text{ in } \B,
 \\
 \bP \nv &=& \bt \text{ on } \dBN,
 \\
 \big( \p{\iv}{\sed}-c\,\Delta\iv + \p{\iv}{\diss} \big)\,\dot{\iv} &=& 0 \text{ in } \B,
 \\
 \Grad\iv\smpc \nv &=& 0 \text{ on } \dB,
 \\
 \dot\iv &\geq& 0 \text{ in } \B.
\end{array}
\label{e:equation_set}
\ee
Here, $\Delta(\bullet)$ denotes the Laplace operator and $\bP\defeq\p{\bF}{\sed}= (1-D(\iv))\,\p{\bF}{\sed_0}$ denotes the first Piola-Kirchhoff stress tensor and $\nv$ denotes the normal vector corresponding to the surface of the body in reference configuration. 
Equations \eqref{e:equation_set}${}_3$ and \eqref{e:equation_set}${}_5$ can be written as Karush-Kuhn-Tucker conditions \cite{KuhTuc:1951:np} (applied to gradient damage e.g. in \cite{LieSteBen:2001:tac,DimHac:2008:amf,WafPolMenBla:2013:age}) 
\eb
 \yieldf \leq 0  ,\quad  \dot\alpha \geq 0 \quad \text{and}\quad  \yieldf\, \dot\iv = 0 \qquad \text{with} \quad
 \yieldf(\bF,\iv,\Grad\iv) \defeq -\p{\iv}{\sed}+c\, \Delta\iv - \p{\iv}{\diss},
 \label{e:kkt_origin}
\ee
for the solution of optimization problems with inequality constraints.
Herein, $\Phi \le 0$ serves as damage evolution criterion, where $\Phi = 0$ corresponds to the case where damage evolves. 
 \section{Mixed finite element formulation \label{s:fem}}
This section introduces a mixed finite element formulation, in which the damage evolution criterion is incorporated via Lagrange multiplier method. 
For this, in the following subsection \ref{ss:continuous_formulation} the continuous formulation and corresponding solution spaces are given.
Based thereon subsection \ref{ss:discretization} presents a corresponding suitable finite element discretization scheme and algorithmic treatment. 

\subsection{Continuous formulation \label{ss:continuous_formulation}}
In what follows, we consider a time-incremental setting, where $[t_n,t]$ is one time interval and the index $n$ corresponds to quantities evaluated at the last time step.
The no-healing constraint  $\dot{\iv} \geq 0$ \eqref{e:equation_set}${}_5$ becomes $\iv-\iv_n \geq 0$.
We define the following history variable
\eb
\bar{\iv}\defeq
\begin{cases}
\iv_n &\quad \text{for no damage evolution} \\ 
\iv   &\quad \text{for damage evolution} 
\end{cases}
\label{e:def_history_variable_continuous}
\ee
and introduce the Lagrangian
\eb
L(\bu,\iv,\lag) = \int_{\B} \Big[ \sed(\bF,\iv) + \frac{\nlpar}{2} \Grad\iv \scp \Grad\iv +\diss(\iv) + \lag\, (\iv - \bar{\iv})\Big] \dX + \varPi^{\mrm{ext}},
\label{e:lagrangian_proposed}
\ee
where $\lag$ is the Lagrange multiplier variable. 
The treatment of the no-healing constraint is based on the following Karush-Kuhn Tucker conditions:
\eb
\lag \leq 0  ,\quad  \iv-\bar{\iv} \geq 0 \quad \text{and}\quad  \lag\, (\iv-{\bar{\iv}}) = 0, 
\label{e:kkt_proposed}
\ee
in which numerical (trial)-values of 
$\lag$ (resulting from iterations of corresponding numerical solution procedures) can be used to identify the evolution and no-evolution cases of \eqref{e:def_history_variable_continuous}:
If the first case of \eqref{e:def_history_variable_continuous} is identified, with $\lag\,(\iv-\iv_n)$ damage evolution is supressed, wheareas in the second case with $\lag\,(\iv-\iv)$ the constraint term vanishes.
Further details with respect to the numerical treatment of conditions \eqref{e:def_history_variable_continuous} and \eqref{e:kkt_proposed} are given in subsection \ref{ss:algorithmic_treatment}.
%
%
The continuous solution variables are sought in the infinite dimensional solution spaces
\begin{align}
\U         &\defeq\{\bu\in H^1(\B;\R^3): \bu=\barbu \text{ on } \dBD \},
 \label{e:defU}\\
 \mathcal{A}&\defeq\{\iv\in H^1(\B):\textstyle \int_{\B}\iv \dX = \mrm{const.}\} ,
 \label{e:defA} \\
\mathcal{L} &\defeq\{\lag \in L^2(\B)\},
\end{align}
where $\barbu$ is the prescribed displacement function on the Dirichlet boundary $\dBD$ and the fixed volume integral $\int_{\B}\iv \dX$ ensures regularity of corresponding discretized tangent-submatrices, since here $\iv$ has no essential boundary condition.
For the given spacially and temporaly variable quantities $\bbf,\bt,\bu^\star \in L^2(\B)\times L^2(\dBN)\times L^2(\dBD)$
we seek the functions $(\bu,\iv,\lag) \in \U\times \mathcal{A}\times \mathcal{L}$ in the stationary point
\eb
L \Rightarrow \underset{\bu,\iv,\lag}{\operatorname{stat}}.
\ee
%
%
Variation and integration by parts (cf. appendix \ref{app:variation_proposed}) together with \eqref{e:kkt_proposed} yields the following strong form:
\begin{align}
  -\Div\bP&=\bbf \text{ in } \B \label{e:bal_lin_mom_prop},\\
  \bP \nv\  &= \bt \text{ on } \dBN \label{e:cauchy_theorem_prop},\\
  \yieldf(\bF,\iv,\Grad\iv)                           & = \lag \text{ in } \B \label{e:strong_form_lag_equal_yield_prop},\\
  \Grad\iv\smpc \nv                                   & = 0 \text{ on } \dB, \\
  \iv-\bar{\iv}                                        & \ \geq\  0 \text{ in } \B \label{e:strong_form_constraint}.
\end{align}
Equation \eqref{e:strong_form_lag_equal_yield_prop} shows, that the Lagrange multiplier $\lag$ corresponds to the function $\yieldf$.
Thus, \eqref{e:kkt_proposed} can be viewed as analogous condition to \eqref{e:kkt_origin}, where evaluation of $\lag$ corresponds to evaluation of the function $\yieldf$.

\subsection{Discretization\label{ss:discretization}}
Let $\T$ be a tetrahedral finite element triangulation of the computational domain $\B$, where $T\in\T$ is one finite element.
Furthermore, let $\V_T$ and $\mathcal{E}_T$ be the set of vertex and mid-edge nodes and $\M_T$ the mid-volume node of one element.
In the following, $(\bullet)\vert_T$ denotes a function defined on $T$ and defined to be zero in all other elements.
For the interpolation of the displacement field $\bu$  we use the standard interpolation scheme
\eb
\bu^{\mrm{h}}\vert_T =
\sum_{I\in \mathcal{N}_T^{(P_2)}}
\bd_{\mrm{u}}^I N_{\mrm{u}}^I\vert_T 
\qquad \text{and} \qquad
\bF^{\mrm{h}}\vert_T =
\Big( \sum_{I\in \mathcal{N}_T^{(P_2)}}
\bd_{\mrm{u}}^I \otimes \Grad N_{\mrm{u}}^I\vert_T \Big) + \bone ,
\label{e:discr_u}
\ee
where $N_{\mrm{u}}^I\vert_T$ are quadratic Lagrangian nodal (P2) basis functions, $\mathcal{N}_T^{(P_2)}=\V_T\cap\mathcal{E}_T$ is the set of nodes in the P2 discretization, and $\bd_{\mrm{u}}^I$ are the corresponding nodal degrees of freedom.
The interpolation of $\iv$ is given as follows:
\eb
\iv^{\mrm{h}}\vert_T = 
\sum_{I\in \V_T} d_{\mrm{\iv}} N_{\mrm{\iv}}^I\vert_T  + d_{\mrm{\iv}}^{\mrm{B}} N_{\mrm{\iv}}^{\mrm{B}}\vert_T, 
\label{e:discr_iv}
\ee
where $N_{\mrm{\iv}}^I\vert_T$ are piecewise linear nodal basis functions and $d_{\mrm{\iv}}^I$ the corresponding degrees of freedom.
The second term in \eqref{e:discr_iv} denotes the volume bubble enrichment term as used e.g., for the MINI interpolation (cf. \cite{BofBreFor:2013:mfe} and \cite{Bra:2007:fel}), 
which is necessary in order to ensure rank sufficiency of the global tangent matrix (cf. count test of section \ref{sss:count_test}).
The bubble enrichment is described as follows:
Let $\Bxi\defeq\{\xi,\eta,\zeta,\kappa\}$ be the set of tetrahedral (reference) volume coordinates. 
Then $N_{\mrm{\iv}}^{{\mrm{B}}}\vert_T$ is given by $N_{\mrm{\iv}}^{\mrm{B}}\vert_T=256\,\xi\,\eta\,\zeta\,\kappa$ and denotes the quartic basis function corresponding to the mid-volume node $\mathcal{M}_T$.
Consequently, $N_{\mrm{\iv}}^{\mrm{B}}\vert_T$ takes the value $1$ in the element center ($(\xi,\eta,\zeta,\kappa)=1/4(1,1,1,1)$) and the value $0$ on all faces of $T$.
The approximation of the gradient of $\iv^{\mrm{h}}$ reads:
\eb
\Grad\iv^{\mrm{h}}\vert_T = 
\sum_{I\in \V_T} d_{\mrm{\iv}} \Grad N_{\mrm{\iv}}^I\vert_T  + d_{\mrm{\iv}}^{\mrm{B}} \Grad N_{\mrm{\iv}}^{\mrm{B}}\vert_T    
\ee

For the Lagrange multiplier $\lag\in \mathcal{L}$ no element continuity is required, since it has no partial derivatives appearing in \eqref{e:bal_lin_mom_prop}-\eqref{e:strong_form_constraint}. 
Therefore the piecewise constant interpolation
\eb
\lag^{\mrm{h}}\vert_T      = 
d_{\lag     }\vert_T,
\label{e:discr_lag}
\ee
is sufficient, where $d_{\lag}\vert_T$ are internal degrees of freedom and the nodal basis function is $1$.
%
%
The history variable $\bar{\iv}$ is stored at each Gauss point $g$ of each element $T$.
The corresponding set is denoted as follows:
\eb
\mathcal{H}^{\mrm{h}} \defeq \{\bar{\iv}^{\mrm{h}}\vert_{T}^g:T\in\T, g\in \G\vert_T \} ,
\ee
where $\G\vert_T$ is the set of Gauss points of the element $T$.
For the sake of readability, in what follows, the script $(\bullet)\vert_T^g$ is omitted.
Further details with respect to algorithmic treatment of the evolution criterion \eqref{e:def_history_variable_continuous} and corresponding update condition of $\bar{\iv}^{\mrm{h}}$  follow in subsection \ref{ss:algorithmic_treatment}.
%
The integration over the triangulated domain and its surface read
\eb
\int_{\B} (\bullet) \dX = \sum_{T\in\T} \int_{T} (\bullet) \dX \qquad \text{ and } \qquad
\int_{\Gamma} (\bullet) \dA = \sum_{T\in\T} \int_{\p{}T\cap\Gamma} (\bullet) \dA. 
\ee
Inserting \eqref{e:discr_u}-\eqref{e:discr_lag} into \eqref{e:lagrangian_proposed} yields the discrete Lagrangian
\eb
\begin{split}
\boxed{L^{\mrm{h}}\defeq 
\sum_{T\in\T} \Big( \int_{T} \sed(\bF^{\mrm{h}},\iv^{\mrm{h}}) + \frac{\nlpar}{2} \Grad\iv^{\mrm{h}}\scp\Grad\iv^{\mrm{h}}+ \diss(\iv^{\mrm{h}}) + \lag^{\mrm{h}}\, (\iv^{\mrm{h}} - \bar{\iv}^{\mrm{h}}) \dX \Big) + \varPi^{\mrm{ext},h}}\\
\text{with} \qquad
\varPi^{\mrm{ext},h} \defeq - \sum_{T\in\T} \Big( \int_{T}  \bu^{\mrm{h}} \scp \bbf(\bX,t) \dX - \int_{\p{}T\cap\dBN} \bu^{h,\mrm{surf}} \scp \bt(\bX,t) \dA \Big)
\end{split}
\label{e:discrete_lagrangian}
\ee
and $\bu^{h,\mrm{surf}}\in L^2(\dBN)$ being the surface interpolation of the displacements. 
All integrals are numerically evaluated with the four point Gauss quadrature rule.
In the numerical tests of section \ref{s:tests} the proposed formulation is denoted by P2$_{\bu}$-P1B$_{\iv}$-P0$_{\lag}$.

\subsubsection{Matrix Formulation \label{sss:matrix_formulation}}
For a given discretization $\T$ we define the following global vectors of degrees of freedom corresponding to the solution fields
\begin{align}
\mD_{u} & \defeq\operatorname{vec}(\bd_{u}^1\vert\cdots\vert\bd_{u}^{n}) \qquad &\text{with} \quad n =\dim\{\mathcal{N}_T^{(P_2)}:T\in\T\},
\label{e:def_global_dof_u} \\
\mD_{\mrm{\iv}}^{\V} & \defeq(d_{\mrm{\iv}}^1, \cdots, d_{\mrm{\iv}}^{o})^T                     \qquad &\text{with} \quad o =\dim\{\V_T:T\in\T\},
\label{e:def_global_dof_iv_vertex} \\
\mD_{\mrm{\iv}}^{{\mrm{B}}} & \defeq(d_{\mrm{\iv}}^{{\mrm{B}},1}, \cdots, d_{\mrm{\iv}}^{{\mrm{B}},p})^T       \qquad &\text{with} \quad p =\dim\{\M_T:T\in\T\},
\label{e:def_global_dof_iv_internal} \\
\mD_{\mrm{\iv}} & \defeq \mD_{\mrm{\iv}}^{\V} \cap \mD_{\mrm{\iv}}^{{\mrm{B}}}, \qquad & 
\label{e:def_global_dof_iv} \\
\mD_{\lag} & \defeq(d_{\lag}^1,\cdots,d_{\lag}^{p})^T. \qquad &
\label{e:def_global_dof_lag}
\end{align}
Here, $\operatorname{vec}(\bullet)$ denotes the vectorization operator where $(\bullet)$ is some $m\times n$-matrix.
With the global solution vector
\eb
\mD \defeq \mD_{\mrm{u}} \cap \mD_{\mrm{\iv}} \cap \mD_{\lag}
\ee
the stationary point $L^{\mrm{h}}\Rightarrow\underset{\bu^{\mrm{h}},\iv^{\mrm{h}},\lag^{\mrm{h}}}{\operatorname{stat}}$ is the solution of the nonlinear equation system 
\eb
\mR(\mD)\defeq \pp{L^{\mrm{h}}}{\mD}=\bzero.
\label{e:discrete_system_nonlinear}
\ee
At each time interval $[t_n,t]$ the following linearized equality condition corresponding to \eqref{e:discrete_system_nonlinear}
\eb
\mR\vert_i + \mK\vert_i \Delta \mD =\bzero 
\quad \text{with the tangent matrix} \quad \mK\defeq \pp{\mR}{\mD}
\ee
is updated with Newton-Raphson iterations, where $i$ denotes the previous iteration.
Writing the global residual vector and tangent matrix in terms of the indiviudal solution fields yields the notation
\eb
 \mR(\mD_{\mrm{u}},\mD_{\mrm{\iv}},\mD_{\lag}) = 
 \begin{bmatrix}
  \displaystyle \pp{L^{\mrm{h}}}{\mD_{\mrm{u}}} &
  \displaystyle \pp{L^{\mrm{h}}}{\mD_{\mrm{\iv}}}   &  
  \displaystyle \pp{L^{\mrm{h}}}{\mD_{\lag}}
 \end{bmatrix}^T
\label{e:global_residual_vector}
\ee
and
\eb
\mK (\mD_{\mrm{u}},\mD_{\mrm{\iv}},\mD_{\lag}) = 
\begin{bmatrix}
 \displaystyle \ppp{L^{\mrm{h}}}{\mD_{\mrm{u}}^2}                 &    \displaystyle  \pppp{L^{\mrm{h}}}{\mD_{\mrm{u}}}{\mD_{\mrm{\iv}}}       & \bzero \\[0.5cm]
 \displaystyle \pppp{L^{\mrm{h}}}{\mD_{\mrm{\iv}}}{\mD_{\bu}}   &    \displaystyle  \ppp{L^{\mrm{h}}}{\mD_{\mrm{\iv}}^2}             & \displaystyle \pppp{L^{\mrm{h}}}{\mD_{\mrm{\iv}}}{\mD_{\lag}}  \\[0.5cm]
 \bzero                                           &    \displaystyle  \pppp{L^{\mrm{h}}}{\mD_{\lag}}{\mD_{\mrm{\iv}}}  & \bzero 
\end{bmatrix}.
\label{e:global_tangent_matrix}
\ee

Since $\mD_{\mrm{\iv}}^{{\mrm{B}}}$ and $\mD_{\lag}$ are internal degrees of freedom, they can be statically condensed (cf \cite{ArnBre:1985:man,BofBreFor:2013:mfe,Bra:2007:fel}).
Thus, after condensation the number of global equations reduces to $\dim{\mR^{\text{cond}}}=\dim{\mD_{\mrm{u}}}+\dim{\mD_{\mrm{\iv}}^{\V}}$, which is equivalent to the number of gobal equations of a P2$_{\bu}$-P1$_{\iv}$ discretization as in \cite{RieBal:2022:fef}.
Note that, since the condensation procedure constitutes only a rearrangement of the system of equations, the condensed and non-condensed discrete systems of equations remain equivalent.
Therefore, the count test of section \ref{sss:count_test} is not affected by the static condensation procedure.

\subsubsection{Algorithmic Treatment\label{ss:algorithmic_treatment}}
\begin{algorithm}[htb]
\caption{Solution strategy for each time step $[t_n,t]$}
\label{a:solution_strategy}
\begin{algorithmic}[]
\State
\State initialize $\mD_{\mrm{u}}\vert^0=\mD^n_{\mrm{u}}$, $\mD_{\mrm{\iv}}\vert^0=\mD^n_{\mrm{\iv}}$ and $\mD_{\lag}\vert^0=\mD^n_{\lag}$  \Comment{initialize from previous time step}
\State update $\bbf=\bbf\vert_t,\bt=\bt\vert_t,\bu^{{\mrm{B}}}=\bu^{{\mrm{B}}}\vert_t$                                                 \Comment{update time-dependent boundary conditions}
\For{$ i=0,...$}                                                                                      \Comment{Newton-Iterations}
 \State FE-update $\mD^{i+1} = \mD^i$ - $(\mK^i)^{-1}\mR^i$                                     \Comment{solve linear system of equations}
 \For{each Gauss point $g\in \G\vert_T$ in each element $T\in \T$}                             
   \State $\bar{\iv}^{h}\vert^{i+1}=\iv_n^{\mrm{h}}$                                                   \Comment{turn on constraint}                                                                        
      \If{$\lag^{\mrm{h}}\vert^{T,i+1}>0 \wedge i\neq 1$}                               \Comment{check evolution criterion}
        \State $\bar{\iv}^{h}\vert^{i+1}=\iv^{h}\vert^{i+1}$                                    \Comment{switch off constraint}
      \EndIf 
 \EndFor
 \If{$\pnorm{\mD^{i+1} - \mD^i}< tol$}                                                          \Comment{Newton-Iteration exit criterion}
 \State exit
 \EndIf
 \EndFor
\end{algorithmic}
\end{algorithm}
This subsection gives an overview of the overall solution procedure and the algorithmic treatment of the update of the history variable $\bar{\iv}^{\mrm{h}}$ (cf. algorithm \ref{a:solution_strategy}).
For each time step $t \leftarrow t_n$ after the initialization of the solution values from the previous time step $n$ and update of the boundary conditions, the algorithm enters the Newton-Loop.
Here, for each iteration $i+1\leftarrow i$ the global linearized system of equations is solved for the nodal solution vector $\mD^{i+1}$.
After the global update the set of history variables is updated.
At each integration point the update of the history parameter $\bar{\iv}^{\mrm{h}}\vert^{i+1}$ corresponding to condition \eqref{e:def_history_variable_continuous} reads:
\eb
\bar{\iv}^{\mrm{h}}\vert^{i+1}\leftarrow
\begin{cases}
\iv^{\mrm{h}}_n            &\quad\text{if }\lag^{\mrm{h}}\vert^{i+1} \leq 0      \quad \text{(no evolution: constraint switched on)} \\
\iv^{\mrm{h}}\vert^{i+1}   &\quad\text{if }\lag^{\mrm{h}}\vert^{i+1} > 0         \quad \text{(evolution: constraint switched off)} 
\end{cases}
\label{e:update_bariv_algorithmic_treatment}
\ee
Consequently, in the first case, with $\lag^{\mrm{h}}\vert^{i+1}(\iv^{\mrm{h}}\vert^{i+1}-\iv^{\mrm{h}}_n)$ the constraint is switched on and evolution is supressed in the upcoming iteration.
In the second case, with $\lag^{\mrm{h}}\vert^{i+1}(\iv^{\mrm{h}}\vert^{i+1}-\iv^{\mrm{h}}\vert^{i+1})=0$ the constraint is switched off.
In the upcoming iteration the value of the Lagrange multiplier remains unchanged and the value of the damage variable can evolve.
Note, that for $i=0$ with $\lag^{\mrm{h}}\vert^0(\iv^{\mrm{h}}\vert^0-\bar{\iv}^{\mrm{h}}\vert^0)=\lag^{\mrm{h}}_n\,(\iv^{\mrm{h}}_n-\iv_n^{\mrm{h}})$ the constraint term is always switched off.
To enable reactivation of the constraint term (which becomes relevant in the case of transition from damage loading to de-loading), for $i=1$ the internal variable is always updated with the first case of \eqref{e:update_bariv_algorithmic_treatment}.
Thus, at an example integration point with the example of damage loading in the previous step ($\lag^{\mrm{h}}_n > 0$ at step $n$), the update sequence reads
\eb
 \lag_n^{\mrm{h}} > 0 \underset{\text{constraint off}}{\overset{i=0}{\rightarrow}}
\lag^{\mrm{h}}\vert^1=\lag^{\mrm{h}}_n
\underset{\text{constraint on}}{\overset{i=1}{\rightarrow}} 
 \begin{cases}
 \lag^{\mrm{h}}\vert^2 \leq 0  \ \text{\scriptsize (de-loading, constraint on)} &   \overset{i=3}{\rightarrow}\cdots \\[0.3cm]
 \lag^{\mrm{h}}\vert^2 > 0     \ \text{\scriptsize (further loading, constraint off)}    &  \overset{i=3}{\rightarrow}\cdots,
 \end{cases}
\ee
where the first case depicts the transition to the de-loading state, while the second case depicts further damage loading.
Note that, since with \eqref{e:update_bariv_algorithmic_treatment} in both cases the value of the history parameter is assigned with quantities that are given from the global iterative procedure ($\iv^{\mrm{h}}_n$ and $\iv^{\mrm{h}}\vert^{i+1}$) no additional storage space for the history parameter is needed.
Furthermore, while in e.g. \cite{LieSteBen:2001:tac,DimHac:2008:amf,WafPolMenBla:2013:age} additional computational resources are needed for the numerical evaluation of some trial function, here the value of $\lag^h\vert^{i+1}$ is also already given from the global Newton iteration
(yet, without increasing the size of the global system due to the static condensation).

 \section{Numerical tests \label{s:tests}}
In this section the proposed formulation is numerically tested.
The count test of section \ref{sss:count_test} numerically evaluates the stability condition for the tangent matrix on a simple cube geometry.
Numerical results of finite element computations on the plate with hole benchmark problem are given in section \ref{ss:pwh}.
For the following numerical tests the AceGen/AceFEM softwarepackage is used.
The linearized system of equations is solved with the PARDISO solver.
The numerical value of the exit criterion of the Newton iterations of algorithm \ref{a:solution_strategy} is set to $tol=10^{-8}$.

\subsection{Count Test\label{sss:count_test}}
In the following, the necessitiy of the volume bubble enrichment in the discretization of $\iv^h$ (cf. section \ref{ss:discretization}) is shown through evaluation of a simple count test.
\begin{table}
\begin{center}
\includegraphics[width=3.5cm]{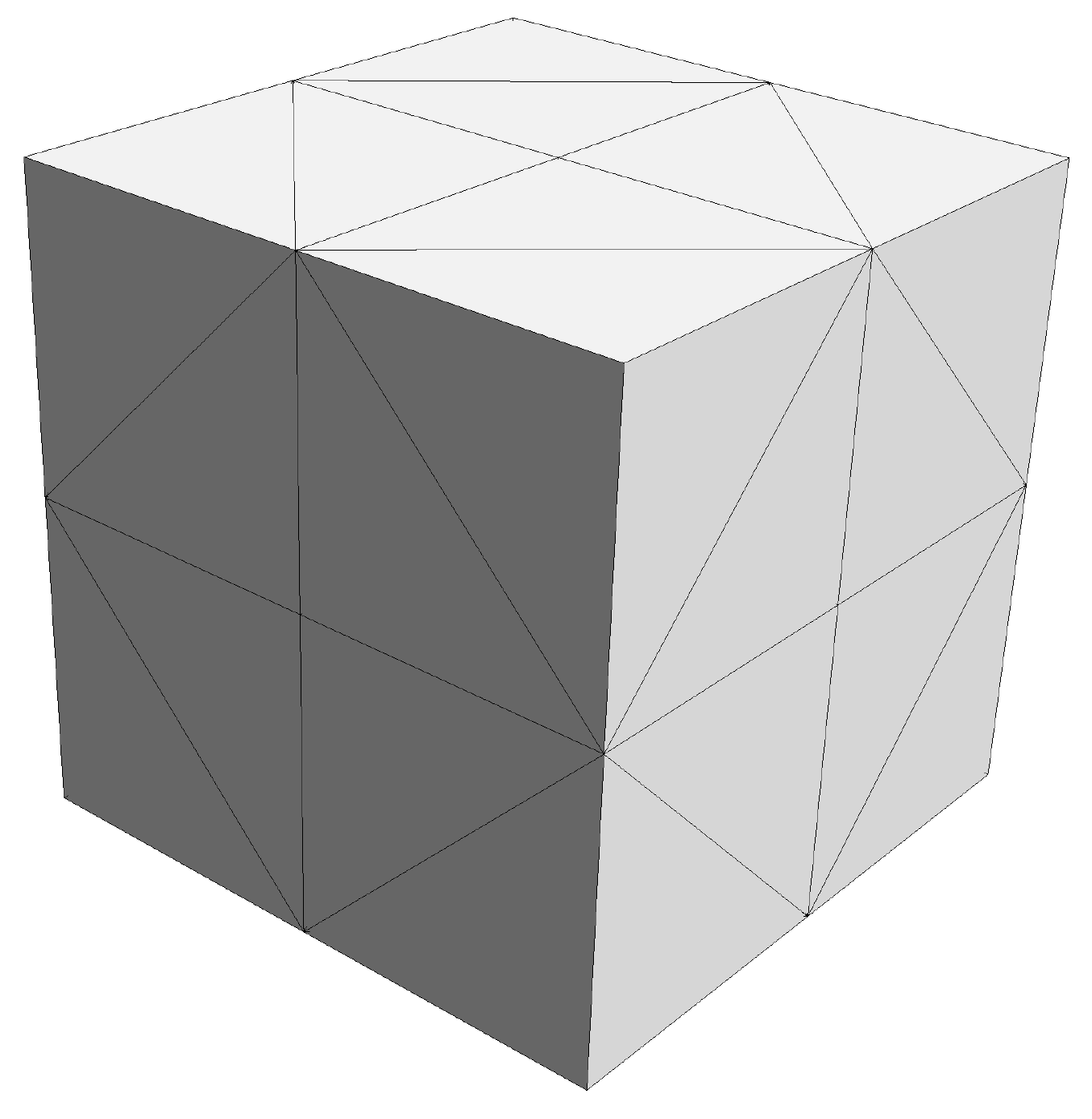}\hspace{0.5cm}
\includegraphics[width=3.5cm]{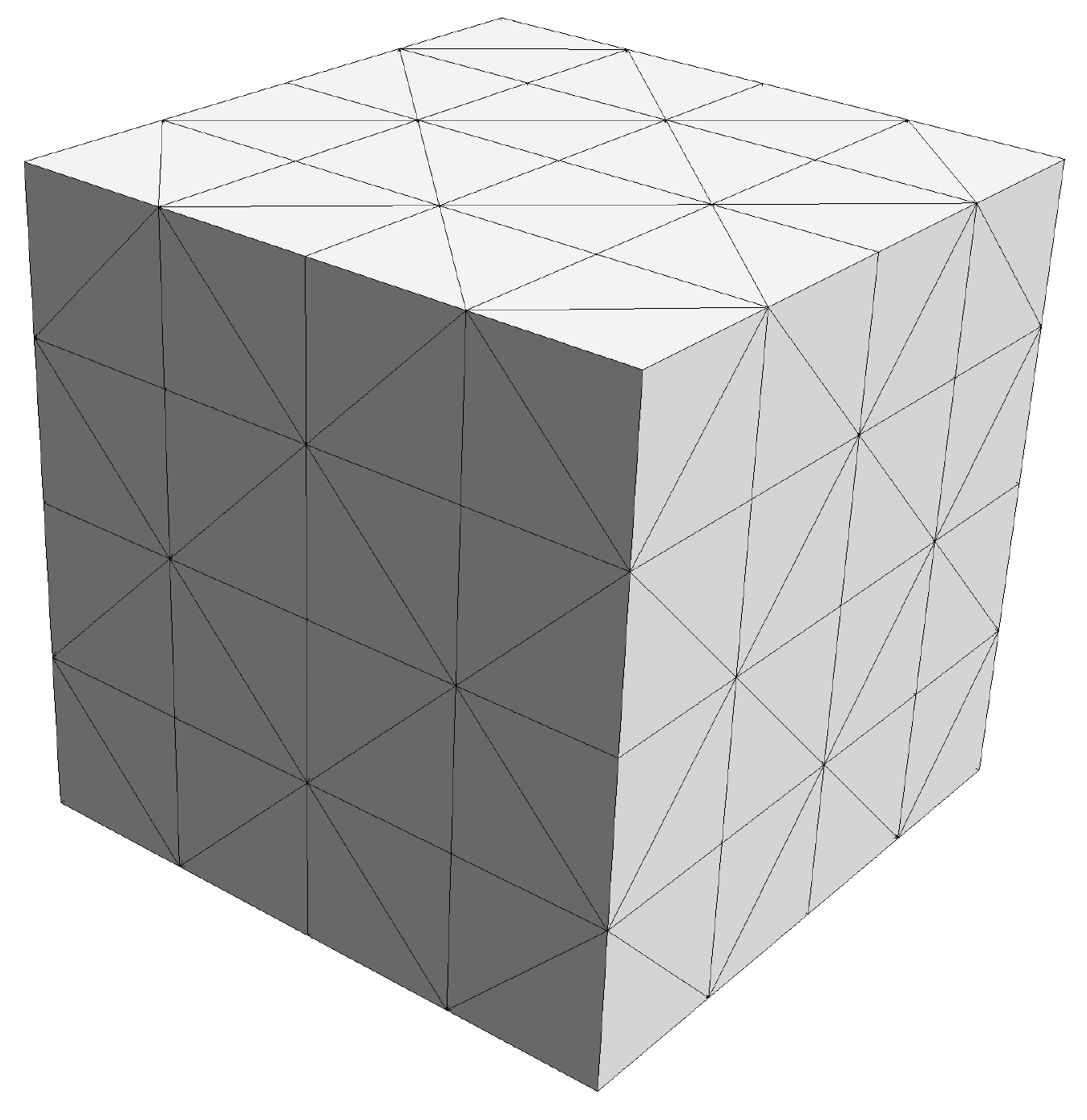}\hspace{0.5cm}
\includegraphics[width=3.5cm]{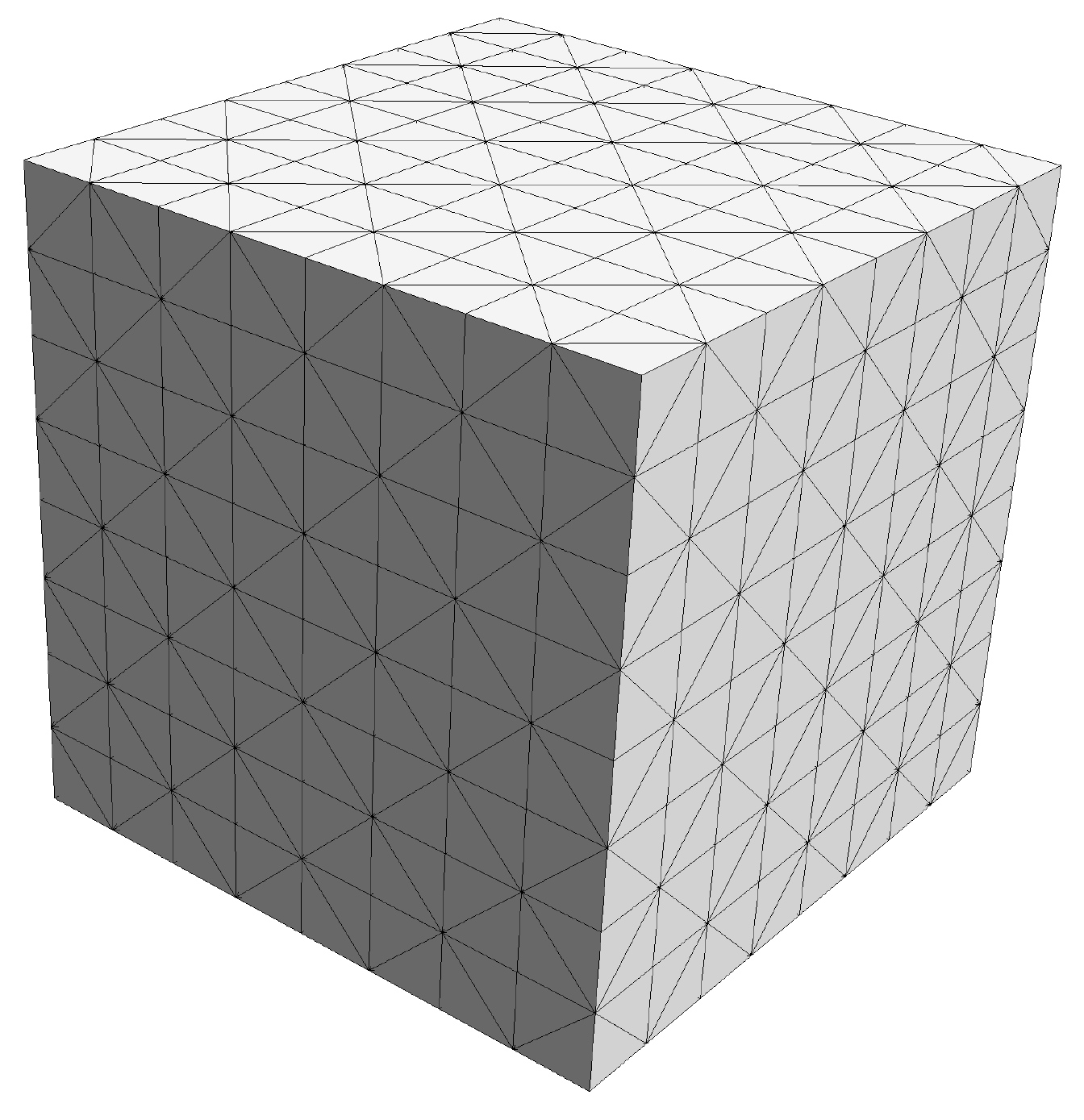}
\begin{tabular}{ccccc} \toprule
Refinement Step &   $\dim{\V}$  & $\dim{\mathcal{M}}$  & Count test (without  $\mD_{\iv}^{{\mrm{B}}}$)  & Count test \eqref{e:count_test} \\ \midrule
1             &   $27$        & $40$                 &     \textcolor{red}{-13}               &   27               \\  
2             &   $125$        & $320$                 &     \textcolor{red}{-195}            &  125               \\ 
3             &   $729$        & $2560$                 &     \textcolor{red}{-1831}          &  729               \\  \bottomrule
\end{tabular}
\end{center}
\caption{Count test  \eqref{e:count_test}  on the unit cube problem with uniform mesh refinement.\label{t:count_test}}
\end{table}
That is, a necessary condition for rank sufficiency of the global tangent matrix is, that the submatrix $\ppp{L^h}{\mD_{\iv}^2}$  of \eqref{e:global_tangent_matrix} must be a full rank matrix and its rank must be greater than the dimension of the submatrix $\ppp{L^h}{\mD_{\lag}^2}=\bzero$.
Thus, the following count test must be fulfilled.
\eb
\dim{\mD_{\iv}} -\dim{\mD_{\lag}} \geq 0\ \footnotemark.
\label{e:count_test}
\ee
\footnotetext{
Another necessary numerical stability condition is the rank sufficiency of the submatrix $\ppp{L^h}{\mD_{\iv}\mD_{\lag}}$.
Note, that in all numerical tests for this submatrix non-zero singular values are observed.
}
To evaluate the count test in a simple example a unit cube geometry with a corresponding structured mesh discretization as shown in the illustration of table \ref{t:count_test} is considered.
For uniform mesh refinement the results of the count test are shown in table \ref{t:count_test}.
It becomes evident that without the enriched degrees of freedom $\mD_{\iv}^{{\mrm{B}}}$ the count test fails.
This is due to the fact, that for each uniform mesh refinement step $s$, the dimension ($\dim\V= (2^s+1)^3$) of the number of vertex nodes is smaller (and grows slower) than the dimension ($\dim\M= 5\times2^{3s}$) of number of mid-element nodes (=number of elements).
On the other hand, if the enriched degrees of freedom $\mD_{\iv}^{{\mrm{B}}}$ are included the count test is passed.
This is due to the fact that, both the number of degrees of freedom $\mD_{\lag}$ of the Lagrange multiplier and the number of enriched degrees of freedom $\mD_{\iv}^{{\mrm{B}}}$ correspond to the total number of elements ($\dim\mD_{\iv}^{{\mrm{B}}}=\dim\mD_{\lag}=\dim{\M}$).
Thus, the difference $\dim\mD_{\iv}-\dim\mD_{\lag}=\dim\mD_{\iv}^{\V}(=\dim\V)$ yields always the number of vertex nodes.
Therefore, by the volume bubble enrichment the count test is always fulfilled and the result of the count test is always $\dim\V$.

\subsection{Plate with Hole Benchmark Problem\label{ss:pwh}}
\begin{figure}
\unitlength1cm
\begin{minipage}[b][][b]{0.38\textwidth}
(a)
 \def\svgwidth{\textwidth}
  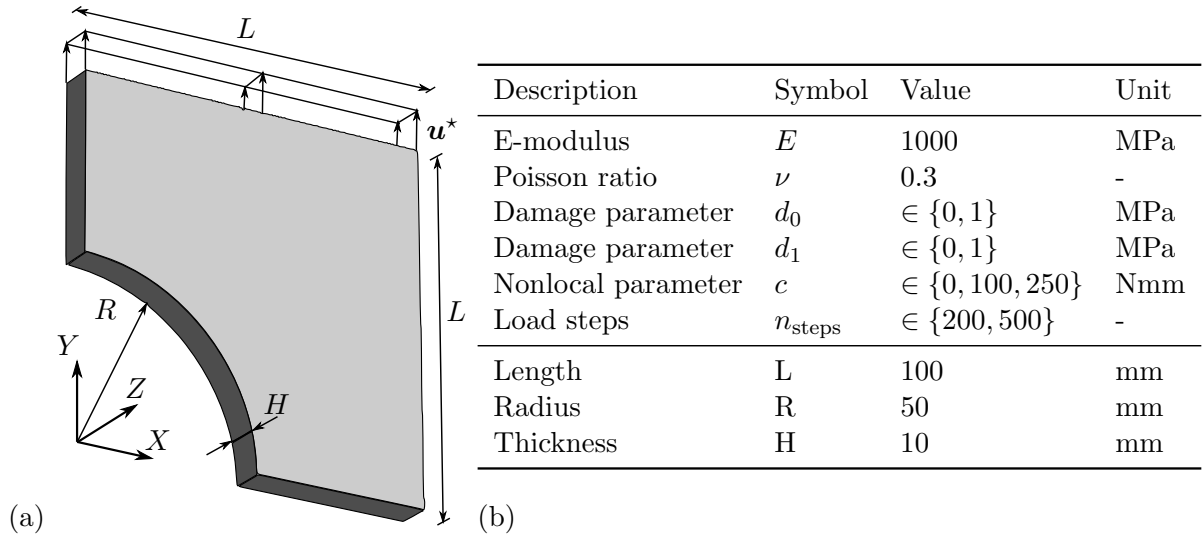
\end{minipage}
\begin{minipage}[b][][b]{0.55\textwidth}
\begin{center}\begin{tabular}{llll} \toprule
Description                    &    Symbol                  & Value         & Unit     \\ 
\midrule               
E-modulus                      &$E $                  &   $1000$           &      MPa           \\  
Poisson ratio                  &$\nu$                 &   $0.3$             &     -     \\  
Damage parameter       &$d_0$                 &   $\in\{0,1\}$               &     MPa   \\  
Damage parameter               &$d_1$                 &   $\in\{0,1\}$      &     MPa   \\  
Nonlocal parameter             &$\nlpar$              &   $\in\{0,100,250\}$    &     Nmm   \\   
Load steps                     &$n_{\mrm{steps}}$     &   $\in\{200,500\}$      &   - \\
\midrule
Length                         & L                    &   $100$             &   mm  \\
Radius                         & R                    &   $50$              &   mm  \\
Thickness                      & H                    &   $10$              &   mm  \\
\bottomrule
\end{tabular}\end{center}
 (b)
\end{minipage}
\caption{
(a) Description of the geometry of the plate with hole benchmark problem. 
(b) Parameters used throughout the tests. 
}
\label{f:pwh_description}
\end{figure}
In this section the proposed formulation is numerically tested on the plate with hole benchmark problem. 
Due to the symmetry only the upper right quarter of the total structure is considered. 
A sketch of the geometry is depicted in  figure \ref{f:pwh_description} (a) and values of the corresponding geometric measures are given in table (b).
Zero displacememts $u_Y=0\ \mrm{mm}$ (in Y-direction) are prescribed at the lower surface $Y=0$ and zero displacements $u_X=0\ \mrm{mm}$ (in X-direction)  are prescribed at the left surface $X=0$.
The problem is displacement-driven with the prescribed displacement $\bu^{\star}=(0, 25, 0)^T \ \mrm{mm}$ at the upper surface $Y=100\ \mrm{mm}$.

\subsubsection{Material model}
For the following computations, we use the damage function
%
\eb
\dv(\iv)\defeq 1-e^{-\iv} 
\label{e:def_sed}
\ee
and the virtually undamaged Neo-Hooke elastic energy density (cf. \cite{Wri:2008:nfe})
\eb
\sed_0 \defeq \frac{\mu}{2}(\Ic-3) + \frac{\lambda}{4}(\JF^2-1)-\frac{\lambda}{2} \ln \JF - \mu \ln \JF
\label{e:neo_hooke_energy}
\ee
with $\Ic=\tr{(\bF^T\smpc \bF)},\ \JF=\det{\bF}$ and the Lam\'e parameters $\lambda=E\nu/((1+\nu)(1-2\nu))$ and $\mu=E/(2(1+\nu))$.
%
%
Meanwhile, the used dissipation function reads
\eb
\diss \defeq \frac{d_1}{2} \iv^2 + d_0 \iv,
\label{e:diss_model}
\ee
where $d_0$ and $d_1$ are damage modeling material parameters. 
Chosing $d_1$ equal to zero corresponds to the model used in \cite{DimHac:2008:amf}, while using both parameters $d_0$ and $d_1$ leads to a model similar to the model used in \cite{WafPolMenBla:2013:age}.
With chosing $d_1=0$  the derivative $\p{\iv}{\diss}=d_0$, which appears in the corresponding update function $\yieldf$ (cf. \eqref{e:kkt_origin}), can also be found in \cite{JunSchJanHac:2019:afa} and \cite{JunRieBal:2021:ent}, where $d_0$ is denoted by $r$ and is called dissipation parameter.
Furthermore, chosing either $d_0$ or $d_1$ equal to zero corresponds to the cases AT1 or AT2 used in \cite{GerDeL:2019:opi}.
An overview of the numerical values of all parameters used throughout the tests can be found in figure \ref{f:pwh_description} (b).

\subsubsection{Force Displacement Curves}
\begin{SCfigure}
\unitlength1cm
\caption{
Mesh-dependent force displacement curves of a local computation ($\nlpar = 0$).
The contourplot corresponds to the refinement and load stage marked with the bullet.
At mesh refinement step 3 ($h_{\mrm{el}}=2.16506 \ \mrm{mm}$) the iterative solution procedure fails to converge at $u_Y^\star\approx 2.5\ \mrm{mm}$ (marked with lightning symbol).
}
\begin{picture}(9,5.1)
\put(0,0){
 \includegraphics[width=0.48\textwidth]{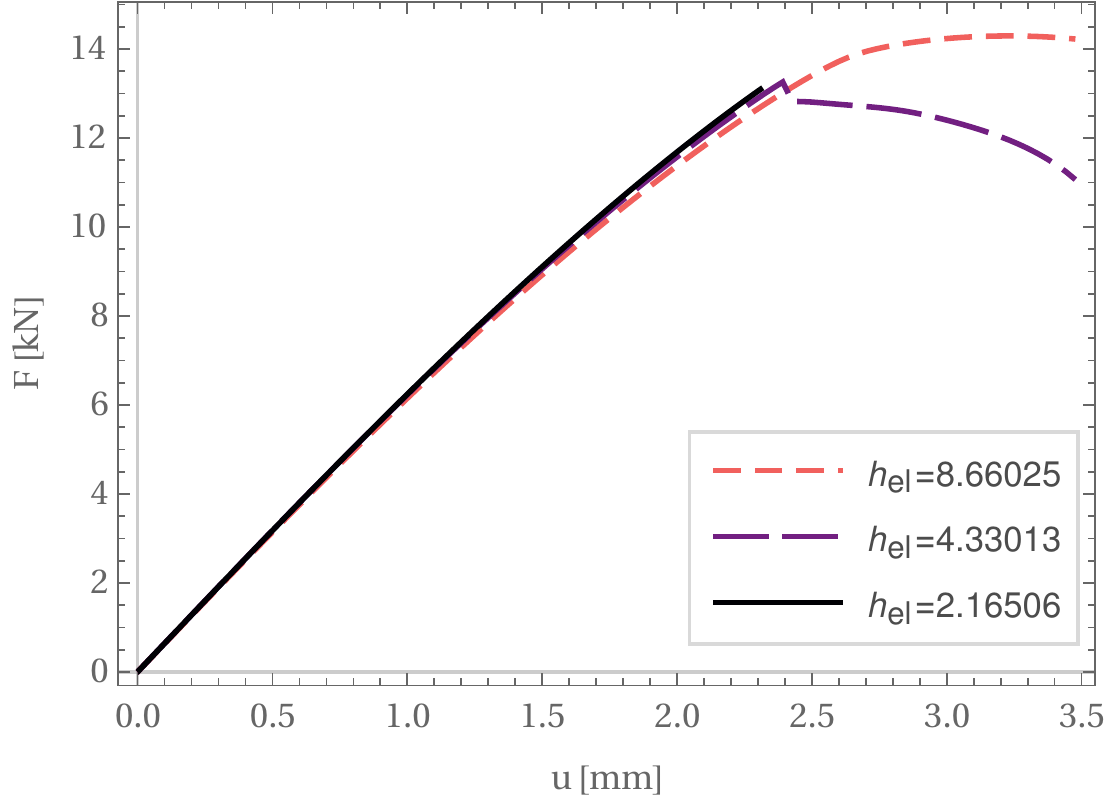}
 \put(-0.4,4.3){$\bullet$}
 \put(-8,4.93){
 \put(5.5,0){\includegraphics[width=0.25cm]{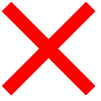}}
   \put(5.6,0.2){\textcolor{red}{\Lightning}}
   }
\put(-5.,1.5){\scalebox{0.7}{
\put(0,0.8){$d_0 = 0$ MPa}
\put(0,0.4){$d_1 = 1$ MPa}
\put(0,0){$c_{\phantom{1}} = 0$ Nmm}
}}
\put(-6.7,3.6){\includegraphics[width=2.5cm]{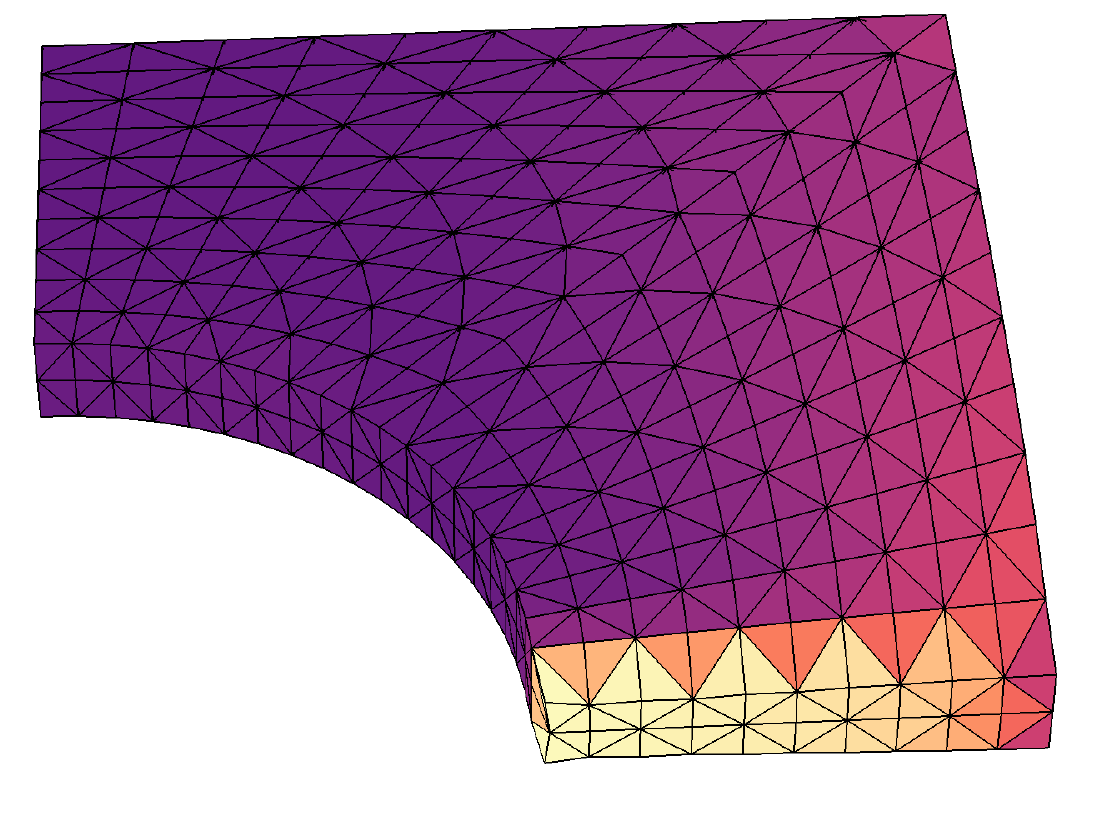}}
}
 \end{picture}
\label{f:pwh_force_displacement_curves_and_contourplots_local}
\end{SCfigure}
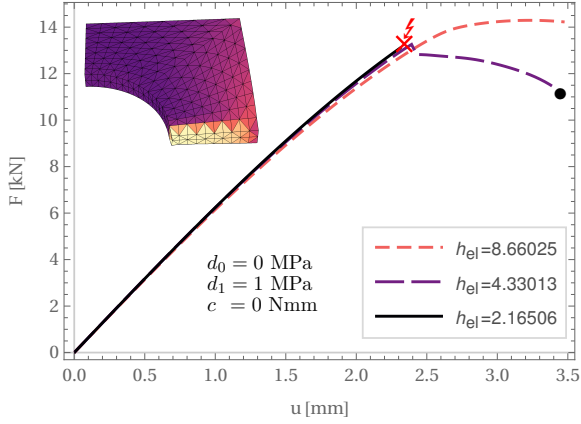
%
%
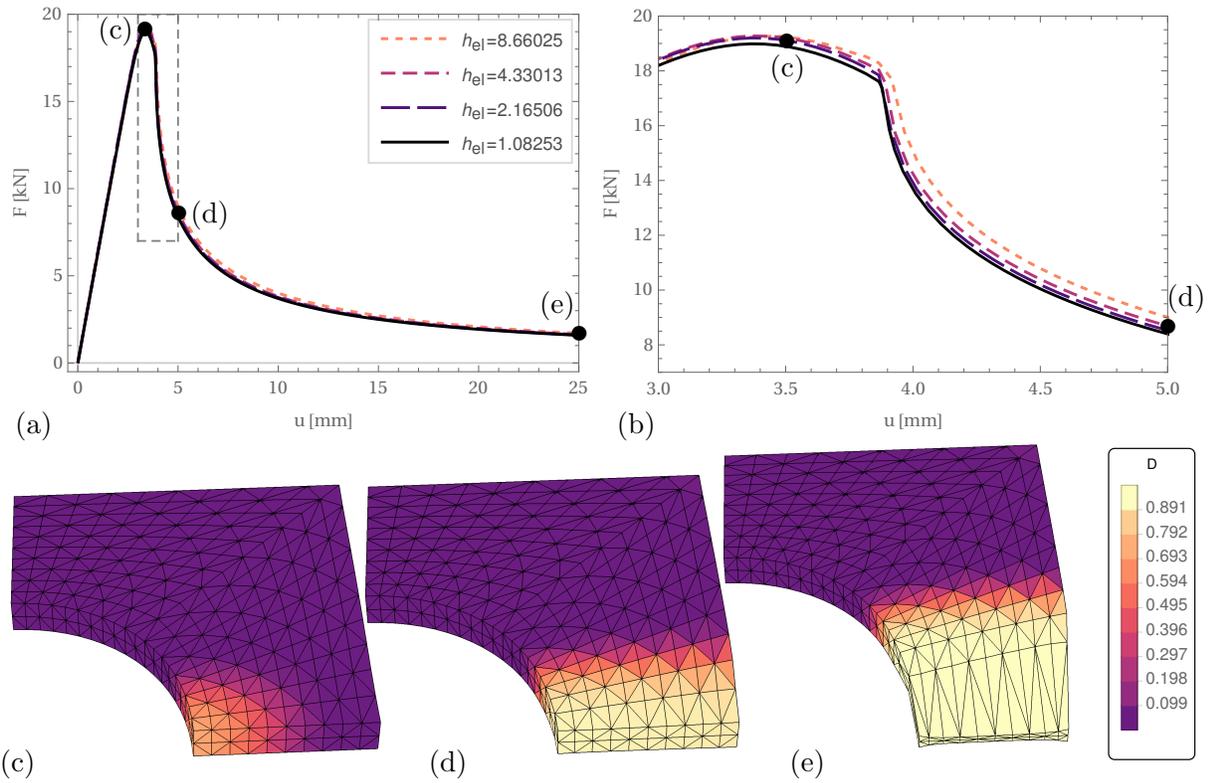
\begin{figure}
\unitlength1cm
\begin{picture}(10,10.2)
\put(0,4.5){
  \includegraphics[width=0.48\textwidth]{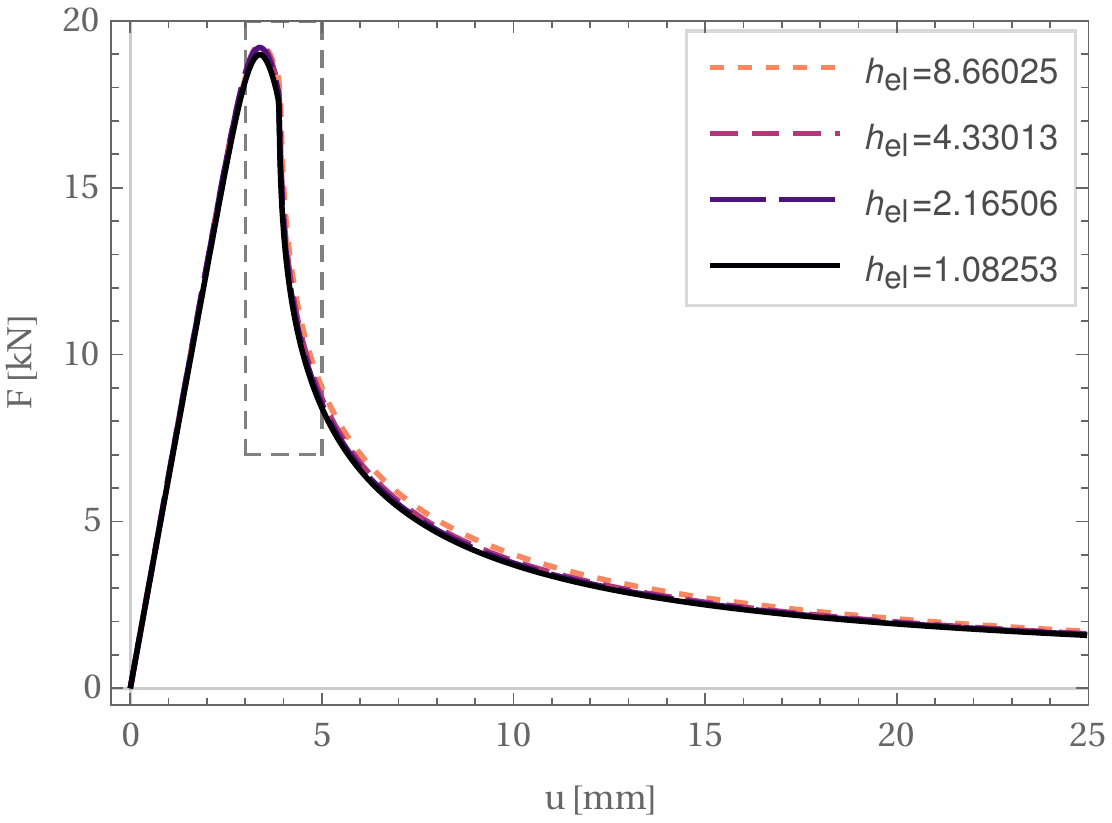}
  \includegraphics[width=0.48\textwidth]{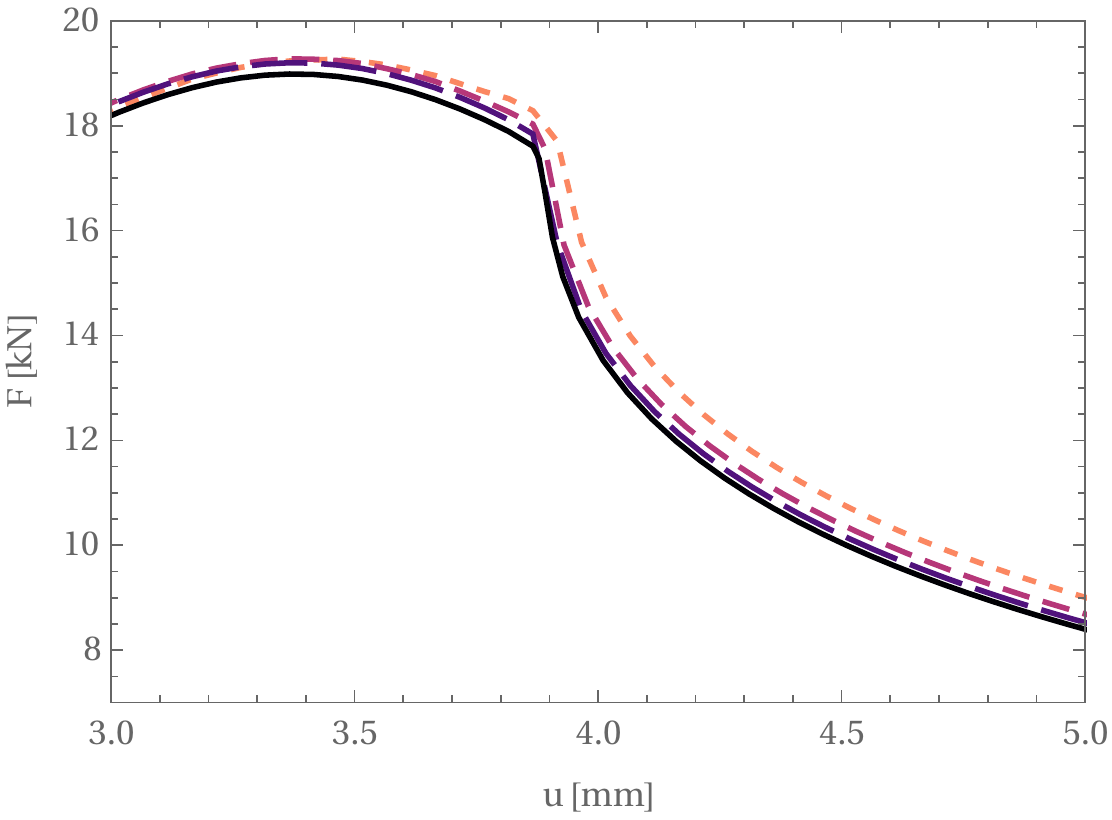}
  \put(-5.37,5.1){
    \put(-0.1,-0.35){(c)}
    \put(0,0){\Large$\bullet$}
  }
  \put(-13.9,5.25){
    \put(-0.5,0){(c)}
    \put(0,0){\Large$\bullet$}
  }
  \put(-0.3,1.3){
    \put(0.11,0.4){(d)}
    \put(0,0){\Large$\bullet$}
  }
  \put(-13.45,2.8){
    \put(0.28,0.){(d)}
    \put(0,0){\Large$\bullet$}
  }
  \put(-8.13,1.2){
    \put(-0.4,0.4){(e)}
    \put(0,0){\Large$\bullet$}
  }
\put(-15.5,0){(a)}
\put(-7.5,0){(b)}
}
\put(0,0){
\put(0,0){\includegraphics[width=5.3cm]{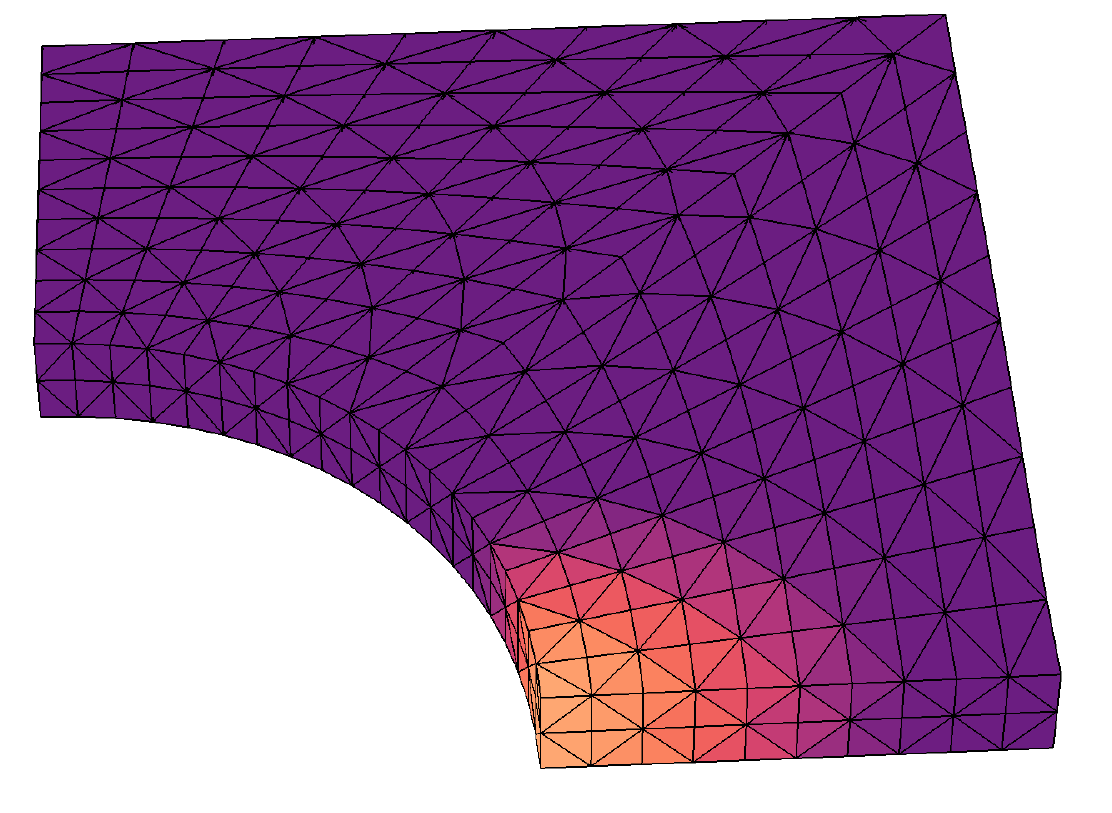}}
 \put(4.7,0){\includegraphics[width=5.4cm]{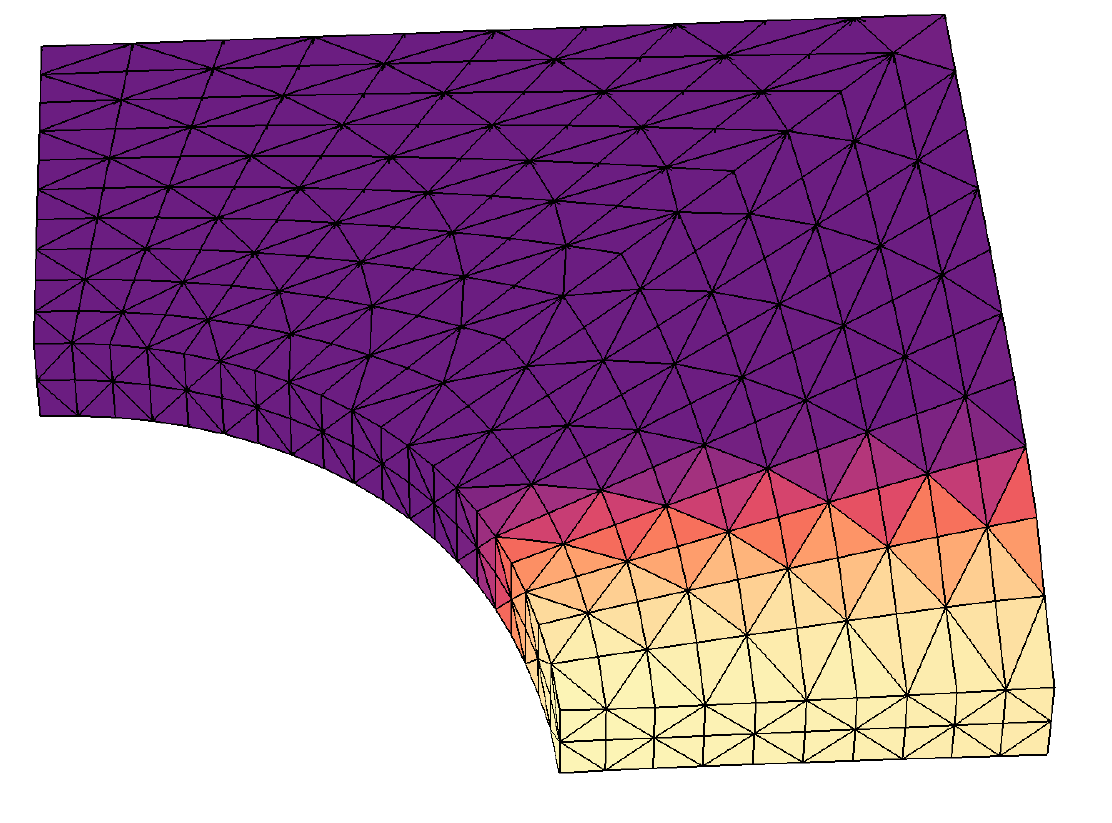}}
 \put(9.5,0){\includegraphics[width=6.4cm]{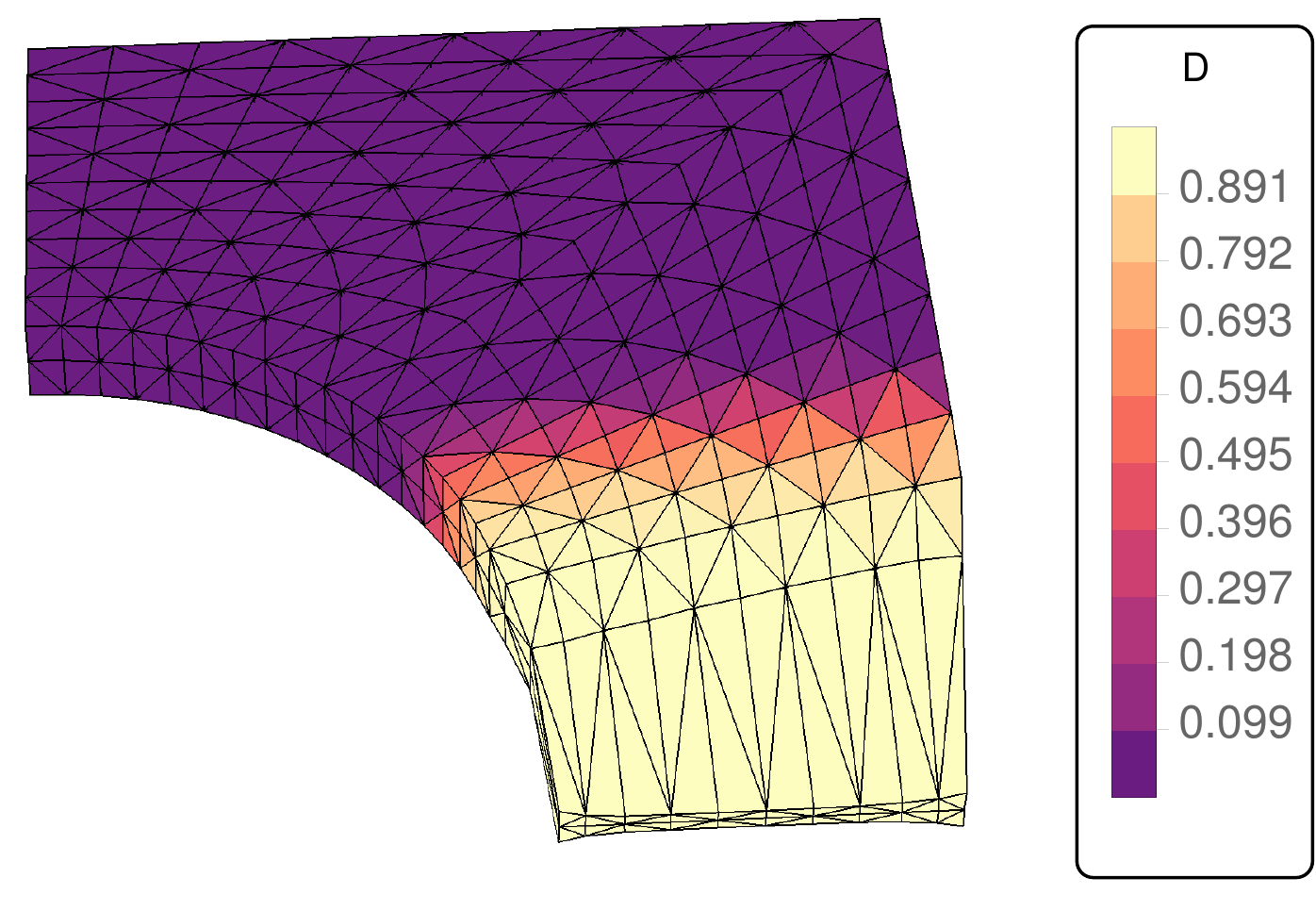}}
 \put(0,0){(c)}
 \put(5.7,0){(d)}
 \put(10.5,0){(e)}
 }
 \end{picture}
\caption{
(a): Force displacement curves corresponding to varying mesh refinement stages for $n_{\mrm{step}}=500$ load steps.
(b): Enlargement of transition area framed with dashed line in (a).
(c)-(e): Contourplots visualizing the damage field $\pnorm{D(\iv)}_{L^2(T)}$ corresponding to load stages $u_Y^\star \in \{3.5,5,25\}$ mm (bullet marks in (a) and (b)).
Clearly, the proposed formulation shows mesh-independent results.
}
\label{f:pwh_force_displacement_curves_and_contourplots}
\end{figure}
In this subsection force displacement curves corresponding to uniform mesh refinement computations are shown.
Depicted in all plots is the value of the reaction force $F$ in $Y$-direction (which is recovered from the upper surface $Y=L$) over the value of the prescribed displacement $u_Y^{\star}$.
In figure \ref{f:pwh_force_displacement_curves_and_contourplots_local} the issue of mesh dependency and loss of convergence of the iterative solution procedure when using a purely local formulation is illustrated.
Here, in order to obtain solutions at all the value of the prescribed displacement is reduced to $u_Y^\star=3.5\ \mrm{mm}$ and is applied with $n_{\mrm{steps}}=500$ load steps.
The nonlocal parameter is zero ($c=0$) and the damage parameters are set to $\{d_0,d_1\}=\{0,1\}\ \mrm{MPa}$.
The contourplot of figure \ref{f:pwh_force_displacement_curves_and_contourplots_local} corresponds to the second refinement computation at the final load stage (marked with bullet).
When comparing the force displacement curves corresponding to the various refinement stages, the mesh dependency of the results are clearly visible.
Furthermore, at mesh refinement step 3 ($h_{\mrm{el}}=2.16506 \, \mrm{mm}$) a loss of convergence of the iterative solution procedure (marked with lightning symbol) is observed.

Figure \ref{f:pwh_force_displacement_curves_and_contourplots} (a) shows force displacement curves corresponding to the proposed gradient enhanced formulation ($\nlpar=100\, \mrm{Nmm}$). 
Contourplots of the second mesh refinement step ($h_{el}=4.33013$ mm) corresponding to the various damage evolution stages marked in figure \ref{f:pwh_force_displacement_curves_and_contourplots} (a) are shown in figure \ref{f:pwh_force_displacement_curves_and_contourplots} (c)-(e).
For the given problem setting at the full load stage (e) nearly completely damaged material states are computed with a maximal value $D_{\mrm{max}}=0.998971$ observed at position $\bX=(85., 0., 10.)^T$.
Since in figure \ref{f:pwh_force_displacement_curves_and_contourplots} (a) the force displacement curves almost coincide, in order to visualize the convergence of the curves an enlarged plot of the strain softening regime (c)-(d) is depicted in figure \ref{f:pwh_force_displacement_curves_and_contourplots} (b).

%
%
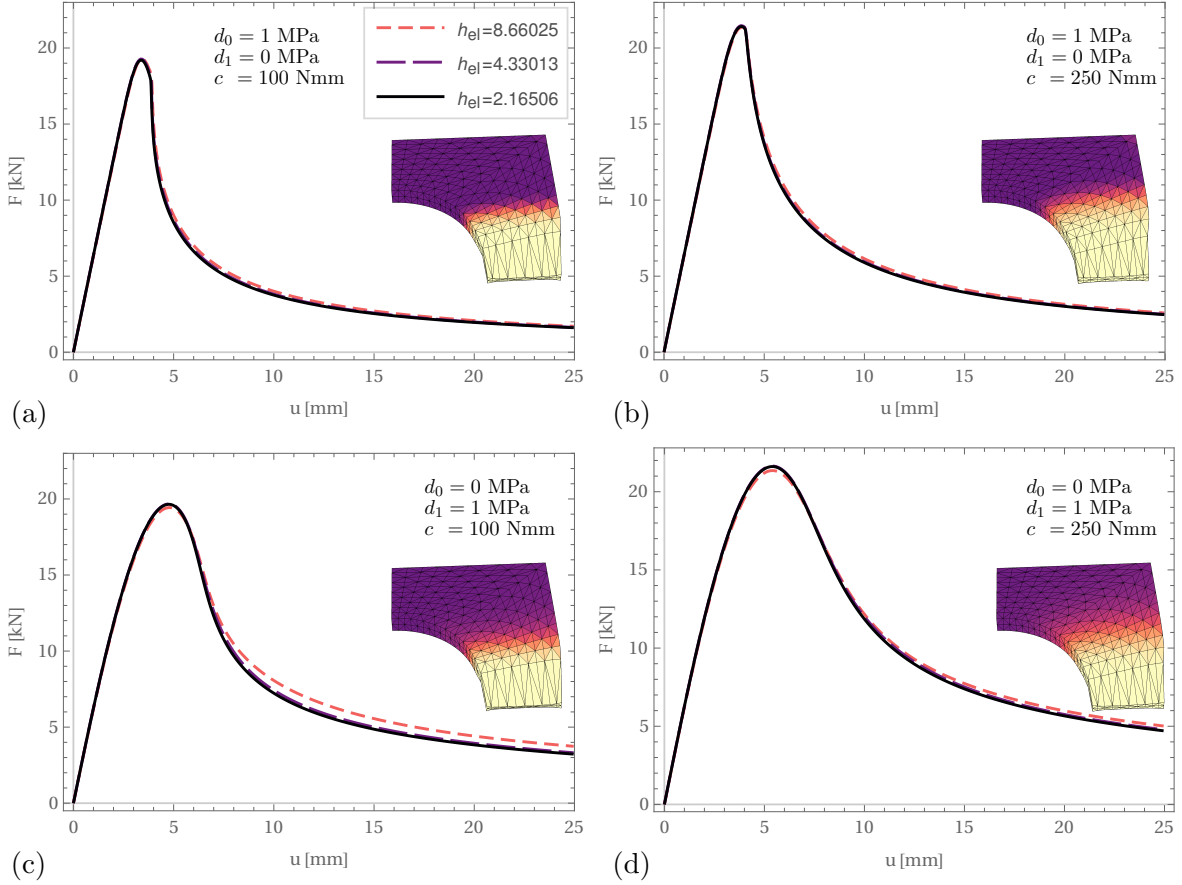
\begin{figure}
\unitlength1cm
\begin{picture}(10,11.7)
 \put(0,6){
 \includegraphics[width=0.48\textwidth]{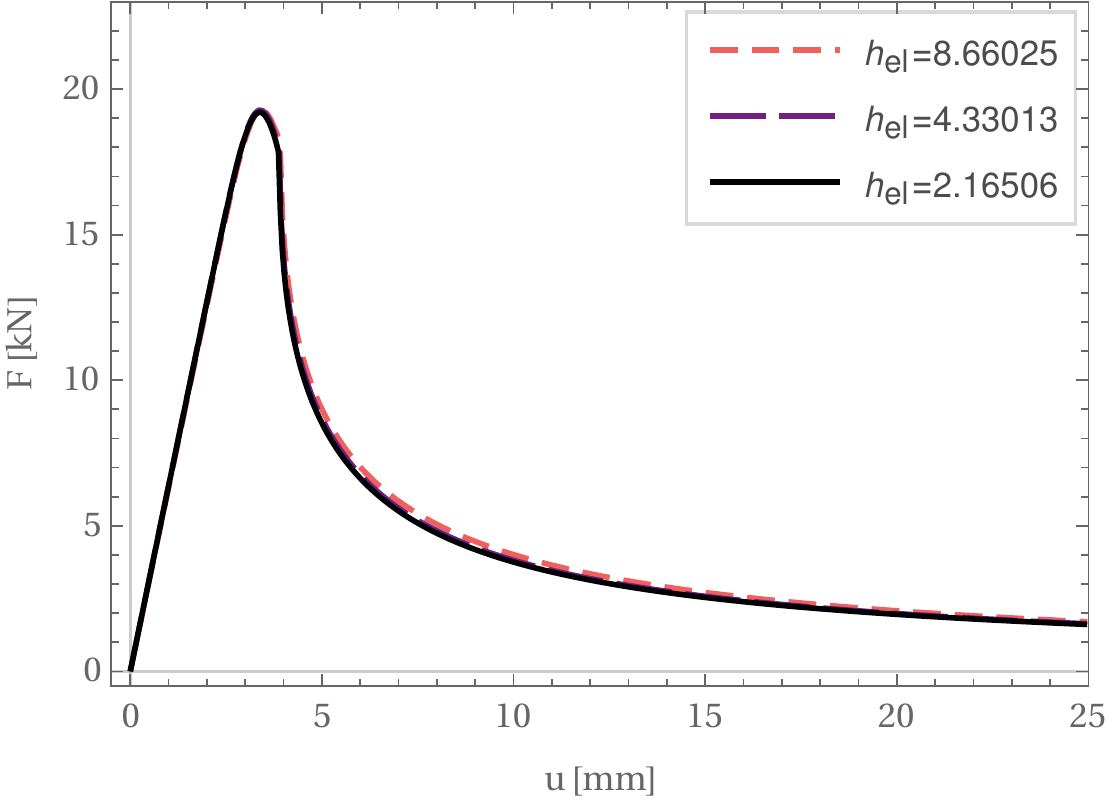}
 \includegraphics[width=0.48\textwidth]{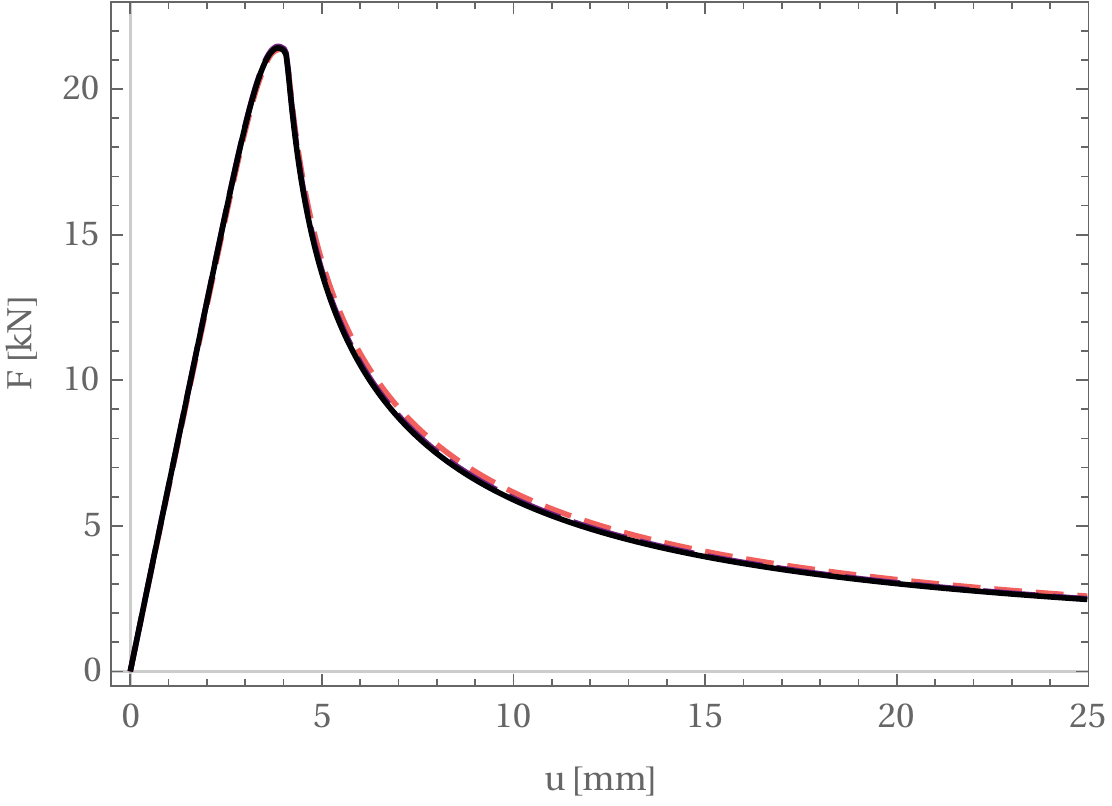}
 \put(-15.5,0){(a)}
\put(-7.5,0){(b)}
\put(-12.8,4.5){\scalebox{0.7}{
\put(0,0.8){$d_0 = 1$ MPa}
\put(0,0.4){$d_1 = 0$ MPa}
\put(0,0){$c_{\phantom{1}} = 100$ Nmm}
}}
\put(-2,4.5){\scalebox{0.7}{
\put(0,0.8){$d_0 = 1$ MPa}
\put(0,0.4){$d_1 = 0$ MPa}
\put(0,0){$c_{\phantom{1}} = 250$ Nmm}
}}
\put(-10.5,1.7){\includegraphics[width=2.5cm]{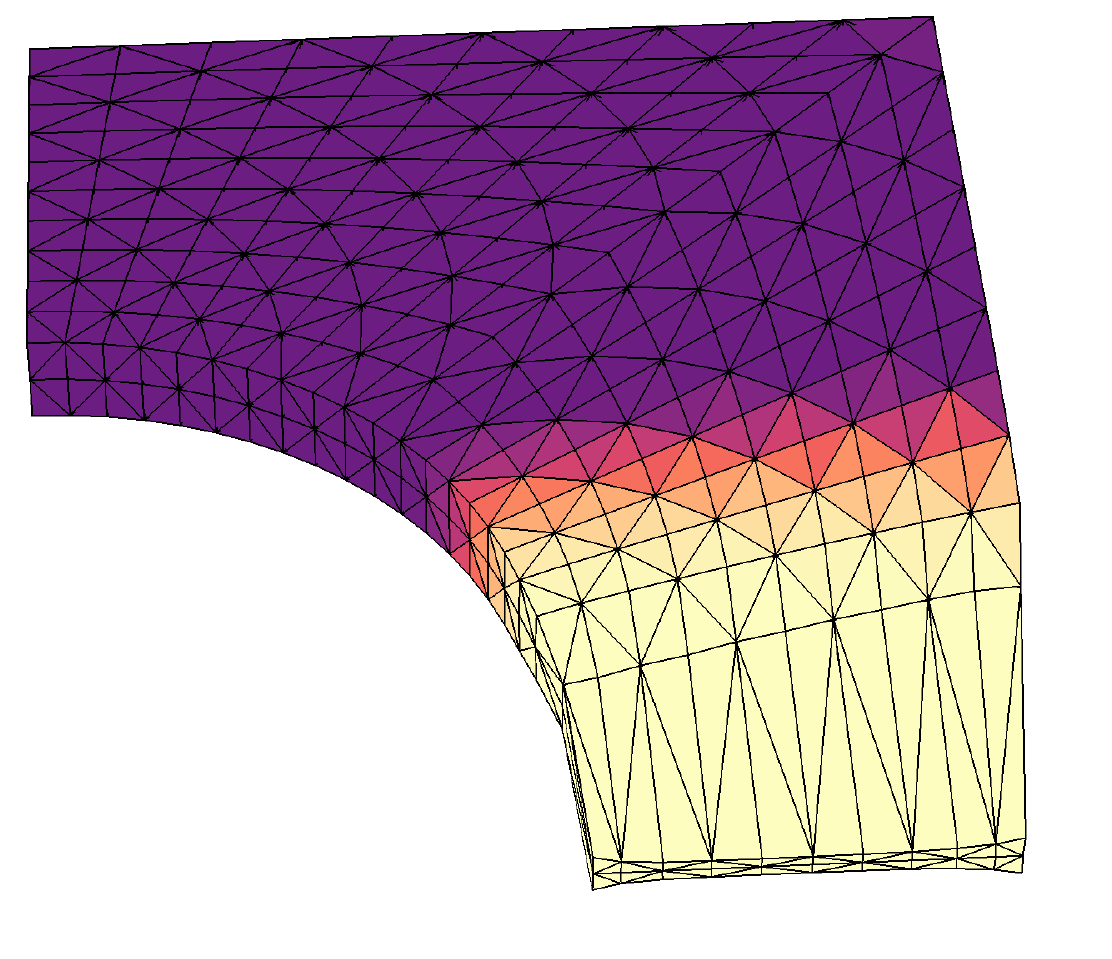}}
\put(-2.65,1.7){\includegraphics[width=2.5cm]{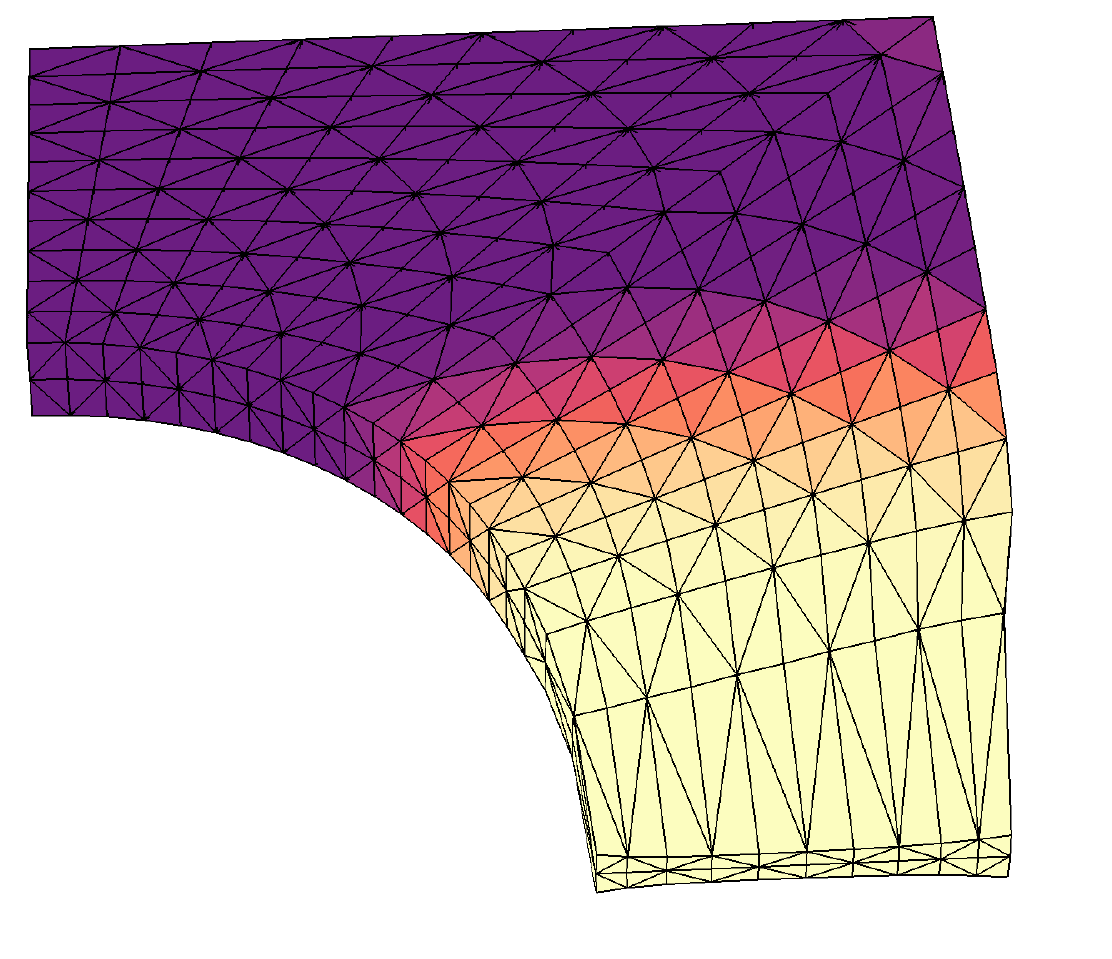}}
}
\put(0,0){
 \includegraphics[width=0.48\textwidth]{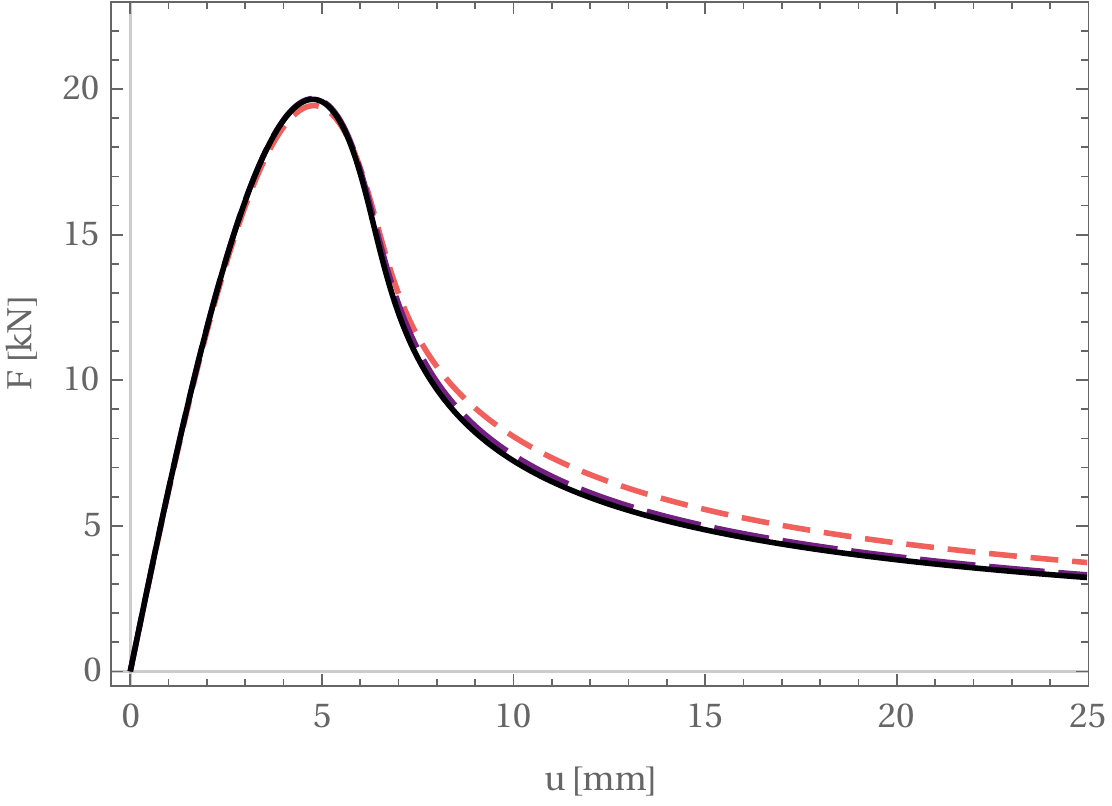}
 \includegraphics[width=0.48\textwidth]{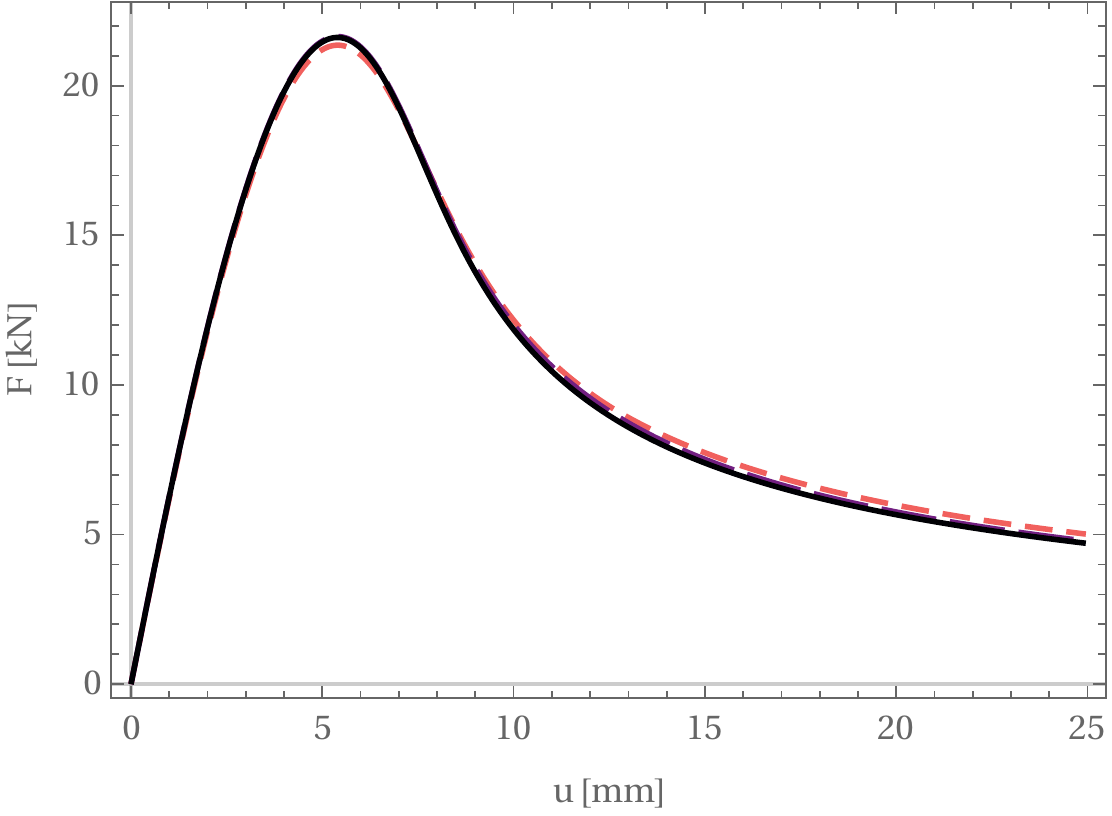}
 \put(-15.5,0){(c)}
\put(-7.5,0){(d)}
\put(-10,4.5){\scalebox{0.7}{
\put(0,0.8){$d_0 = 0$ MPa}
\put(0,0.4){$d_1 = 1$ MPa}
\put(0,0){$c_{\phantom{1}} = 100$ Nmm}
}}
\put(-2,4.5){\scalebox{0.7}{
\put(0,0.8){$d_0 = 0$ MPa}
\put(0,0.4){$d_1 = 1$ MPa}
\put(0,0){$c_{\phantom{1}} = 250$ Nmm}
}}
\put(-10.5,2){\includegraphics[width=2.5cm]{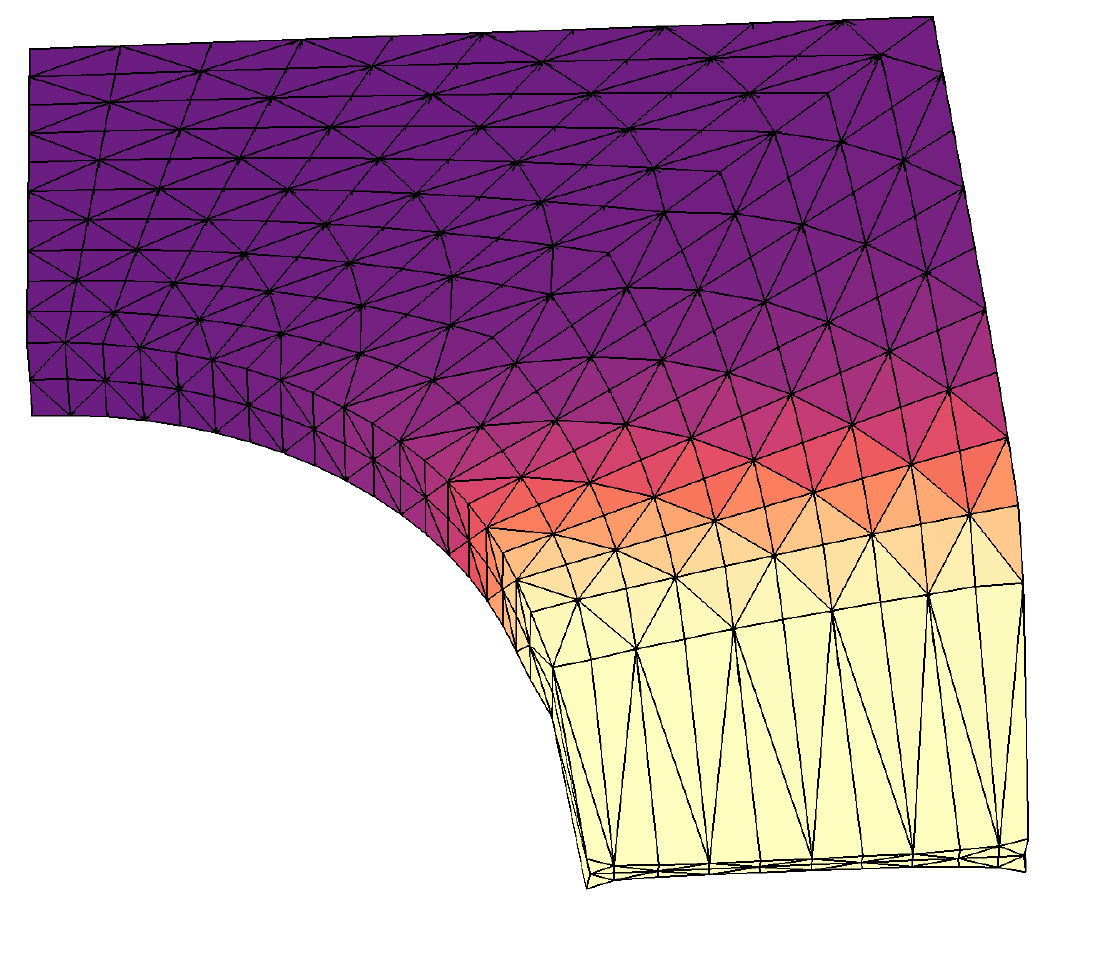}}
\put(-2.45,2){\includegraphics[width=2.5cm]{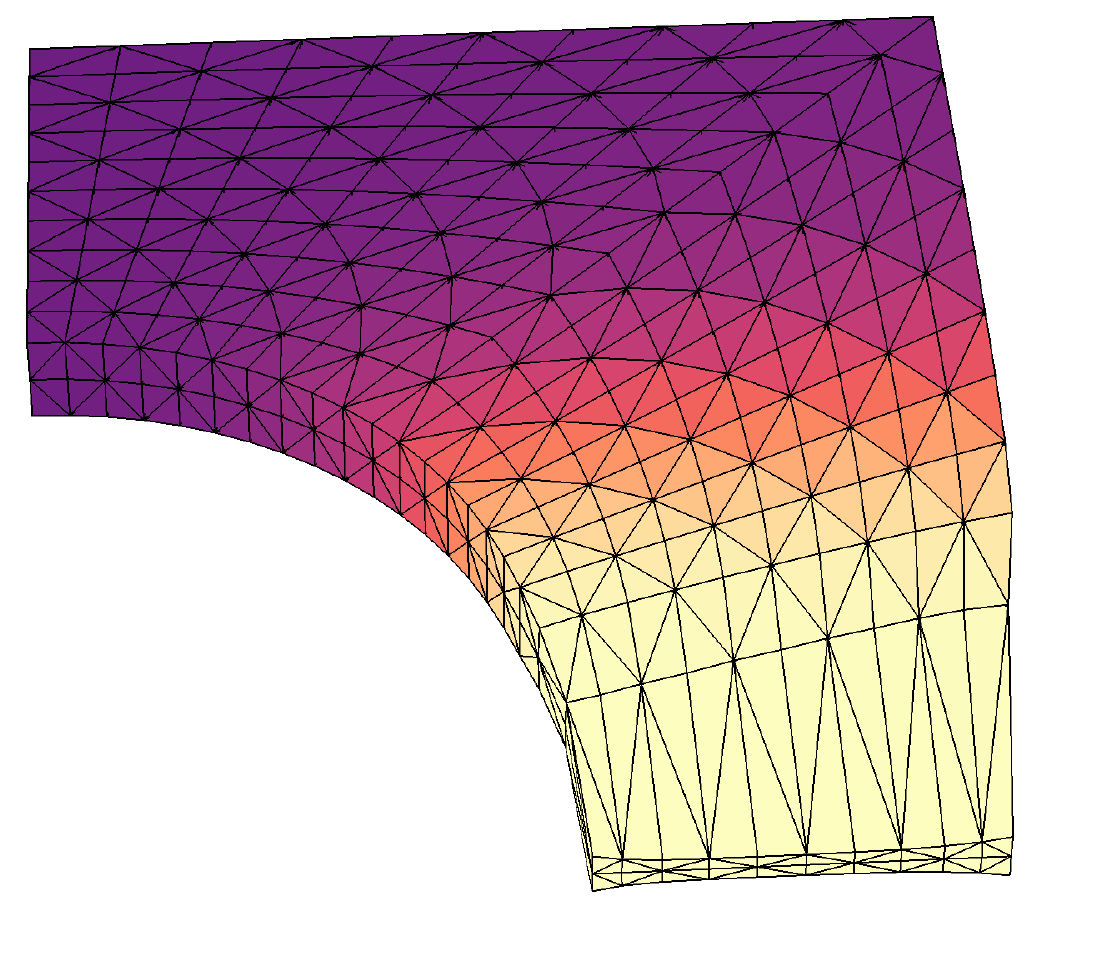}}
}
\end{picture}
\caption{
Force displacement plots for varying parameters and damage contourplots corresponding to the final load stages.
From plots (a) and (c) on the left to the plots (b) and (d) on the right the nonlocal parameter $c$ is increased from $100\ \mrm{Nmm}$ to $250\ \mrm{Nmm}$.
From plots (a) and (b) on the top to plots (c) and (d) on the bottom the value of the damage parameter $d_1$ is changed from $0\ \mrm{MPa}$ to $1\ \mrm{MPa}$.
}
\label{f:pwh_parameter_study}
\end{figure}

In figure \ref{f:pwh_parameter_study} force displacement curves and contourplots analogously to figure \ref{f:pwh_force_displacement_curves_and_contourplots} (a) and (e) are depicted for varying material parameters.
Here, from left (figure \ref{f:pwh_parameter_study} (a),(c)) to right (figure \ref{f:pwh_parameter_study} (b),(d)) the value of the nonlocal parameter $c$ is increased from $c=100\ \mrm{Nmm}$ to $c=250\ \mrm{Nmm}$.
Meanwhile, from top (figure \ref{f:pwh_parameter_study} (a),(b)) to bottom (figure \ref{f:pwh_parameter_study} (c),(d)) the type of dissipation function is changed from $\{d_0,d_1\}=\{1,0\}\ \mrm{MPa}$ to $\{d_0,d_1\}=\{0,1\}\ \mrm{MPa}$.
It becomes evident, that in the computations of figure \ref{f:pwh_force_displacement_curves_and_contourplots} and figure \ref{f:pwh_parameter_study}, mesh independent solutions are obtained. 
Furthermore, even for the present severely damaged states $D_{\mrm{max}}\approx 0.999$ the iterative solution procedure converges.

\subsubsection{Comparative Study}
\begin{figure}
\unitlength1cm
\begin{picture}(10,12.2)
\put(0,6.5){
  \includegraphics[width=0.48\textwidth,height=5.7cm]{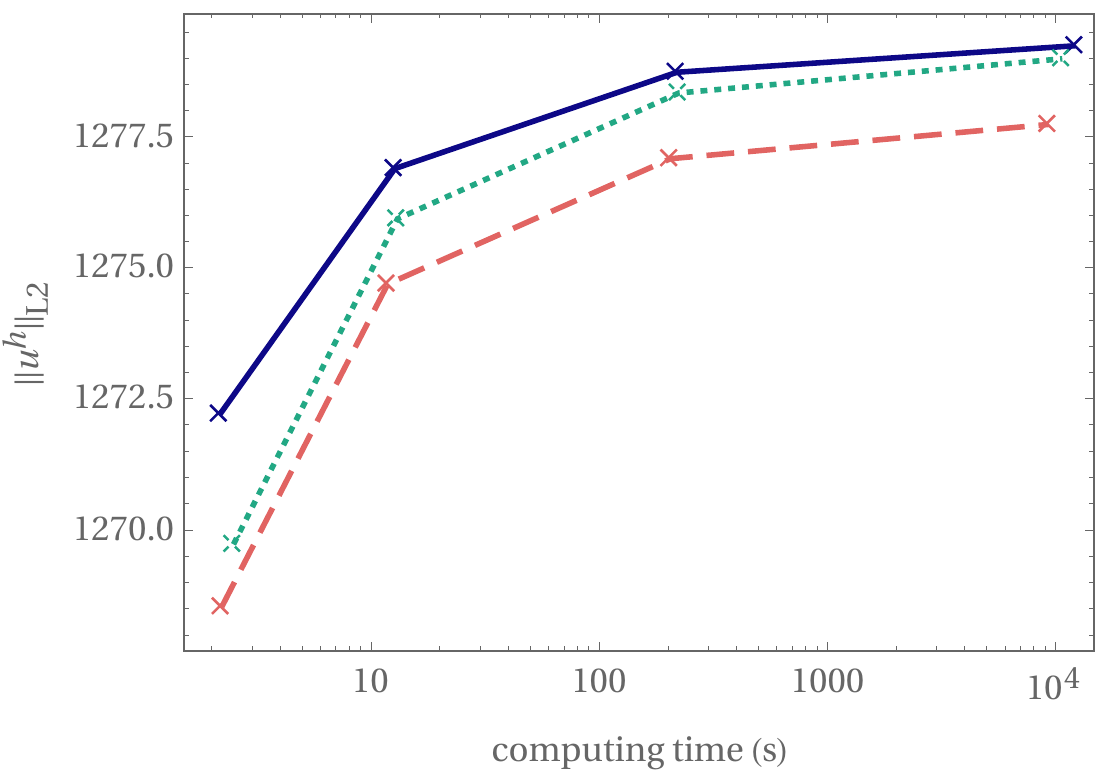}
  \includegraphics[width=0.48\textwidth]{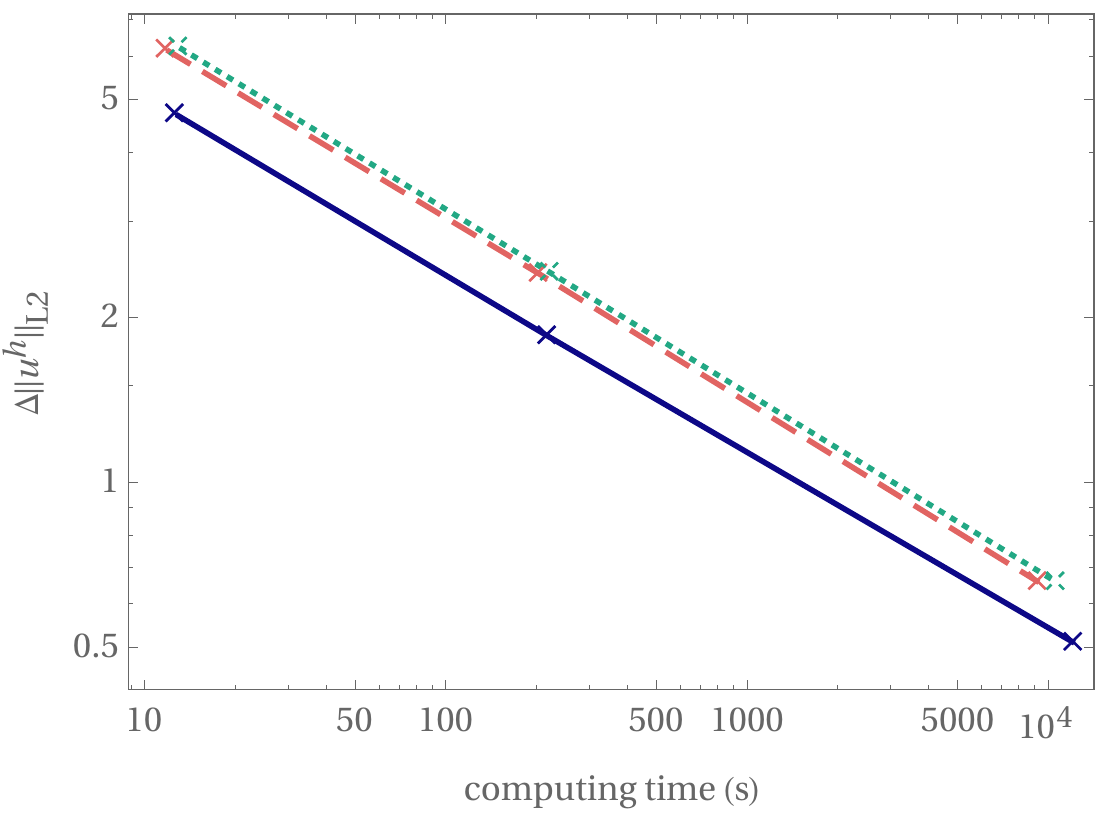}
  \put(-13.5,1.5){\includegraphics[width=5.5cm]{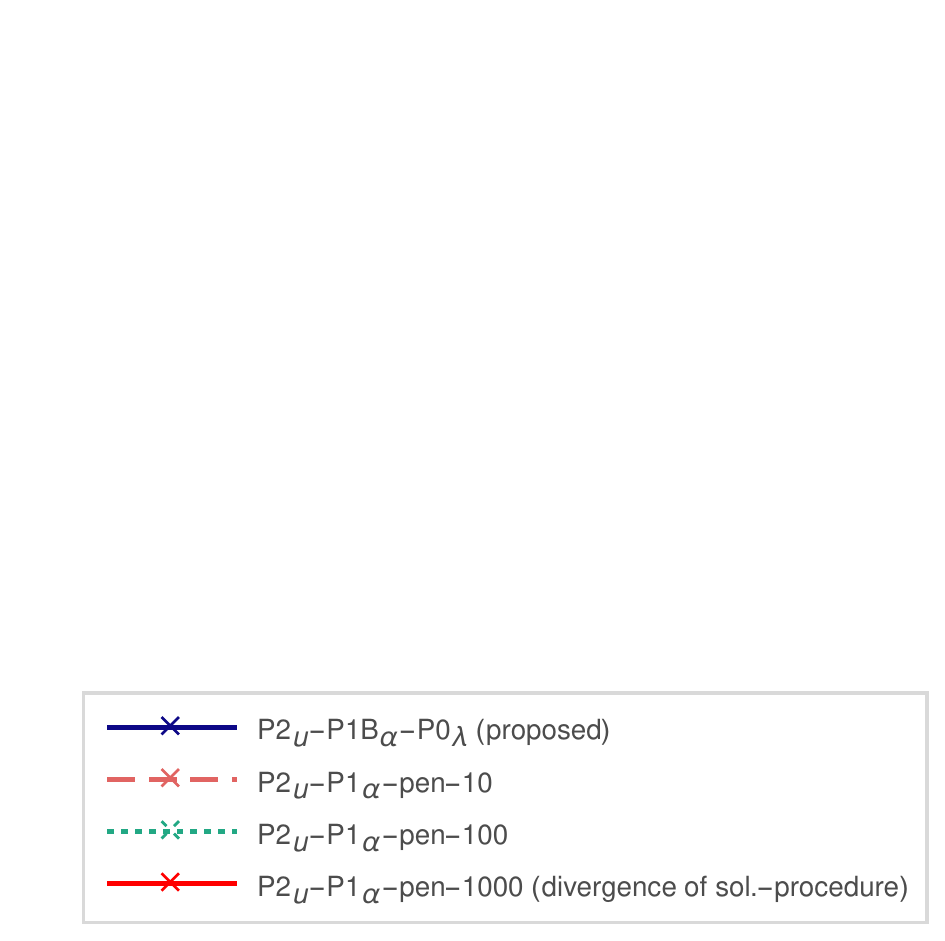}}
  \put(-14,0.92){
   \put(0,0){\includegraphics[width=0.15cm]{fig/pwh_performance_red_x.pdf}}
   \put(0.1,0.2){\textcolor{red}{\Lightning}}
   \put(1,0){\includegraphics[width=0.15cm]{fig/pwh_performance_red_x.pdf}}
   \put(1.1,0.2){\textcolor{red}{\Lightning}}
   \put(2.9,0){\includegraphics[width=0.15cm]{fig/pwh_performance_red_x.pdf}}
   \put(3.0,0.2){\textcolor{red}{\Lightning}}
   \put(5.5,0){\includegraphics[width=0.15cm]{fig/pwh_performance_red_x.pdf}}
   \put(5.6,0.2){\textcolor{red}{\Lightning}}
   \put(5.35,0.4){
		\begin{tikzpicture}
           \draw[color=gray,-{Latex[length=1mm,width=1mm]}] (0,0) -- (-0.1,0.35)
           ;
     \end{tikzpicture} }
  }
  
  \put(-15.5,0){(a)}
\put(-7.5,0){(b)}
}
\put(0.85,0){\includegraphics[width=0.923\textwidth]{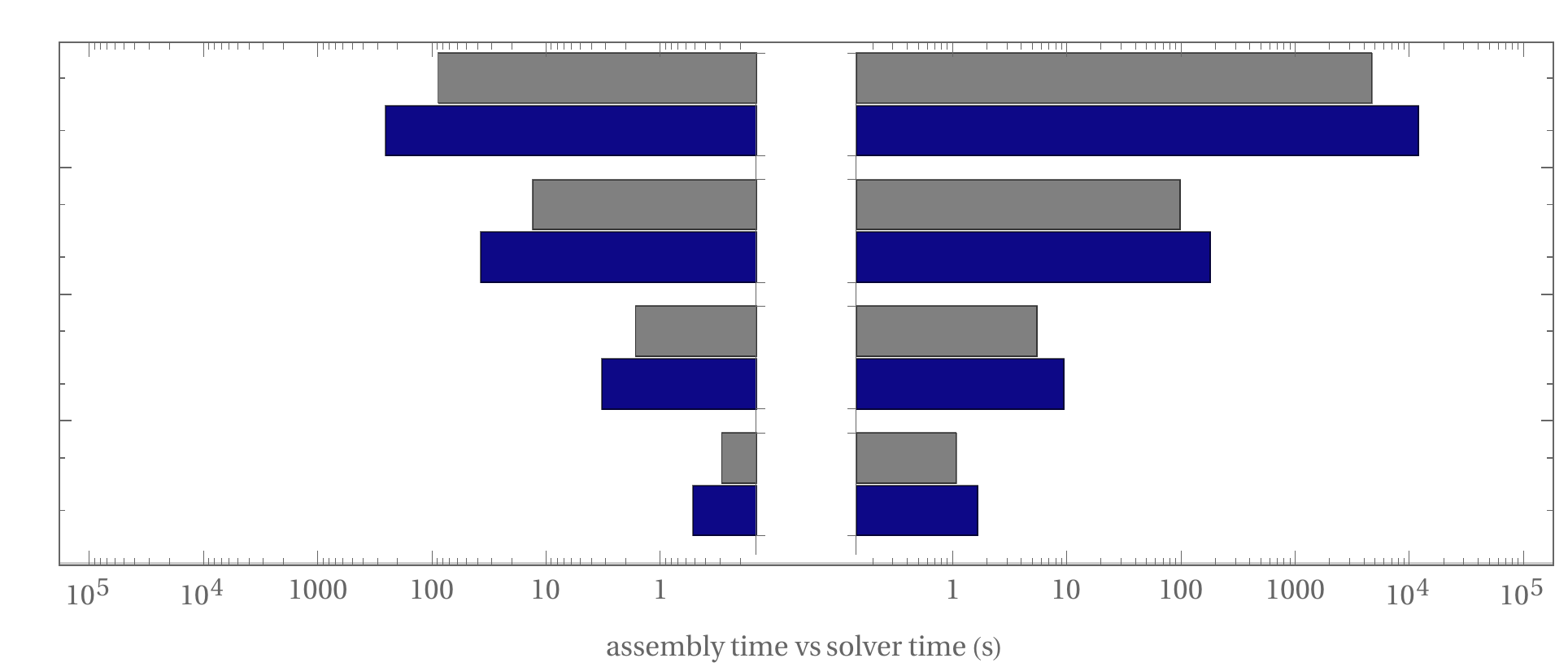}
\put(-14.1,1.1){\includegraphics[width=3.5cm]{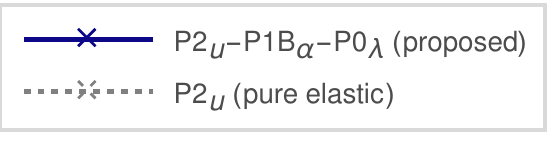}}
\put(-15.5,0){(c)}
}
 \end{picture}
\caption{
Performance plots:
(a): Convergence plot of the $L^2$-norm $\pnorm{\bu^h}_{L^2(\T)}$ of the displacement solution over total computing time for uniform mesh refinenement.
(b): Rate of change $\Delta \pnorm{\bu^h}_{L^2(\T)}$ (cf. \eqref{e:def_rate_of_change}) of the displacement norm over total computing time.
(c): Total assembly times (left) and solver times (right) of the proposed element for each refinement step compared to corresponding times of a purely elastic reference computation.
As can be seen, the proposed mixed finite element formulation converges faster than the competitive formulation.
Yet, the computational effort is not far from a simple elastic simulation with a standard P2-element.
}
\label{f:pwh_comparative_study}
\end{figure}
%
%
%
For the comparative numerical study of this section the proposed formulation is compared to a penalty formulation, which is in line with the approach of \cite{DimHac:2008:amf} and~\cite{WafPolMenBla:2013:age}.
The approach is based on stationarization of the following potential
\eb
P^h\defeq
\sum_{T\in\T} \Big( \int_{T} \sed(\bF^h,\iv^h) + \diss(\bar{\iv}^h,\Grad\iv^h) + \frac{p}{2}\, (\iv^h - \bar{\iv}^h)^2 \dX \Big) + \varPi^{\mrm{ext},h}
\label{e:discrete_lagrangian_comparative}
\ee
where the term $\frac{\penpar}{2}\, (\iv^h - \bar{\iv}^h)^2$ constitues the penalty term.
While the discretization of the displacements is the same as in the proposed approach given in \eqref{e:discr_u}, here the nodal damage variable is discretized with the linear Lagrange shape functions:
\eb
\iv^{h}\vert_T = \sum_{I\in \V_T} d_{\iv} N_{\iv}^I\vert_T
\quad \text{and} \quad
\Grad\iv^h\vert_T = \sum_{I\in \V_T} d_{\iv} \Grad N_{\iv}^I\vert_T  .
\label{e:discr_iv_comparative}
\ee
The internal variable of the comparative formulation is defined as
\eb
\bar{\iv}^h \defeq 
\begin{cases}
 \underset{\bar{\iv}^h}{\operatorname{root}\,}\tilde{\yieldf}(\iv^h,\bar{\iv}^h) \quad&\text{if } \tilde{\yieldf}(\iv^h,\bar{\iv}^h_n) > 0 \\
 \bar{\iv}^h_n                                                                       &\text{else} ,
\end{cases}
\label{e:def_history_variable_comparative}
\ee
where $\tilde{\yieldf}$ is the yield function with the equation $\tilde{\yieldf}=0$ being obtained from the strong form corresponding to the variation of $P^h$ with respect to $\bar{\iv}^h$.
In the case of damage evolution (indicated by $\tilde{\yieldf}(\iv^h,\bar{\iv}^h_n)>0$) the evolved value of $\bar{\iv}^h$ is the root of the equation $\tilde{\yieldf}=0$ and in the other case $\bar{\iv}^h=\bar{\iv}^h_n$ remains unchanged from the previous iteration.
Note, that the formulation corresponding to \eqref{e:discrete_lagrangian_comparative}-\eqref{e:def_history_variable_comparative} differs from \cite{DimHac:2008:amf} and \cite{WafPolMenBla:2013:age} in the sense that here $\sed\defeq\sed(\bF^h,\iv^h)$ is a function of the nodal variable, while in the mentioned literature $\sed\defeq\sed(\bF^h,\bar{\iv}^h)$ is a function of the history variable.
Here, the first approach is used because in numerical tests better numerical robustness of the solution procedure was observed.
Furthermore, $\tilde{\yieldf}$ remains linear in $\bar{\iv}^h$ so that in the update procedure of $\bar{\iv}^h$, at the integration points, no additional Newton sub-iterations are necessary for finding $\operatorname{root} \tilde{\yieldf}$.
The computations corresponding to the comparative penalty formulation are denoted by P2$_{\bu}$-P1$_{\iv}$-pen``$X$'' where ``$X$'' is the numerical value of the penalty parameter $p$.

The convergence behavior of the proposed formulation compared to the penalty formulation is shown in the plots of figure \ref{f:pwh_comparative_study} (a) and (b).
The shown results correspond to computations with uniform mesh refinement on the plate with hole benchmark problem with $u^{\star}_Y=5\ \mrm{mm}$ (cf. figure \ref{f:pwh_force_displacement_curves_and_contourplots} (d)).
Here, the number of load steps is $n_{\mrm{steps}}=100$.
For the comparative formulation results with a penalty value of $p=10$ and $p=100$ are shown.
Note, that for $p \geq 1000$ the iterative solution procedure fails to converge.
From the results of figure \ref{f:pwh_comparative_study} (a), where the $L^2$-norm of the displacements is plotted over the computing time corresponding to each refinement step, it can be observed, that the formulations converge.
Furthermore, as the value of the penalty parameter increases, the results of the penalty formulation approach the results of the proposed formulation.
In order to compare the rates of convergence, the plot of figure \ref{f:pwh_comparative_study} (b) shows the rate of change
\eb
\Delta\pnorm{\bu^h}_{L^2} \defeq \pnorm{\bu^h}_{L^2}^{(s+1)}-\pnorm{\bu^h}_{L^2}^{(s)}
\label{e:def_rate_of_change}
\ee
of the $L^2$-norm between the mesh refinement steps $s$ and $s+1$ plotted over the computing times.
For increasing mesh refinement a converging rate of change $\Delta\pnorm{\bu^h}_{L^2}$ for both formulations becomes visible.
However, by comparing the curves of figure \ref{f:pwh_comparative_study} (b) an improved convergence behavior of the proposed P2$_{\bu}$-P1B$_{\iv}$-P0$_{\lag}$ formulation compared to the penalty approaches becomes evident.
The diagram in figure \ref{f:pwh_comparative_study} (c) visualizes the total assembly times (left) and linear solver times (right) of the proposed P2$_{\bu}$-P1B$_{\iv}$-P0$_{\lag}$ formulation compared to a purely elastic reference computation with a simple P2 elastic displacement element.
It becomes evident, that the computational effort of the proposed damage formulation is not far from the purely elastic standard P2 formulation.

\subsubsection{Behavior Under Cyclic Loading}
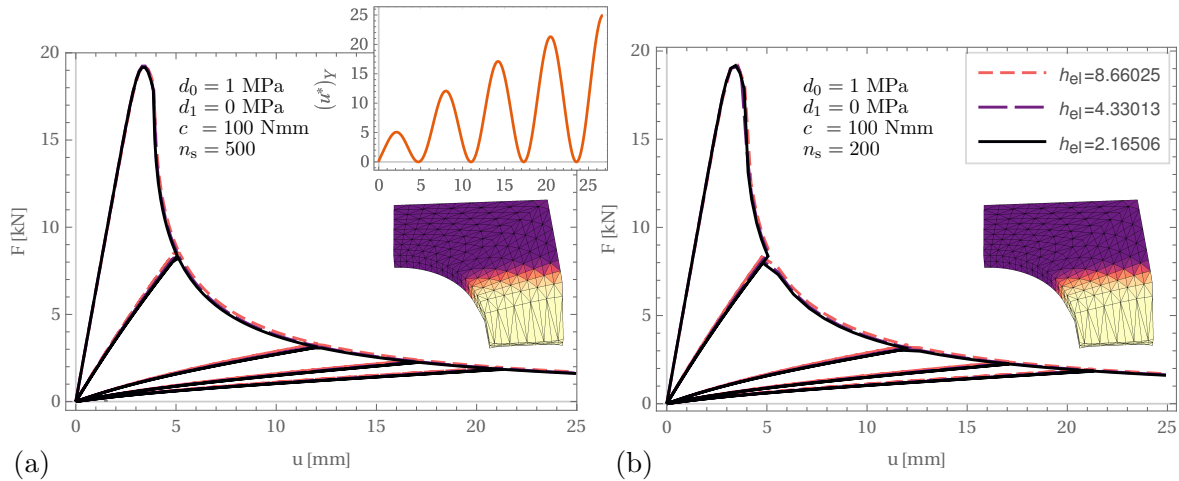
\begin{figure}
\unitlength1cm
\begin{picture}(10,6.2)
 \put(0,0){
 \includegraphics[width=0.48\textwidth]{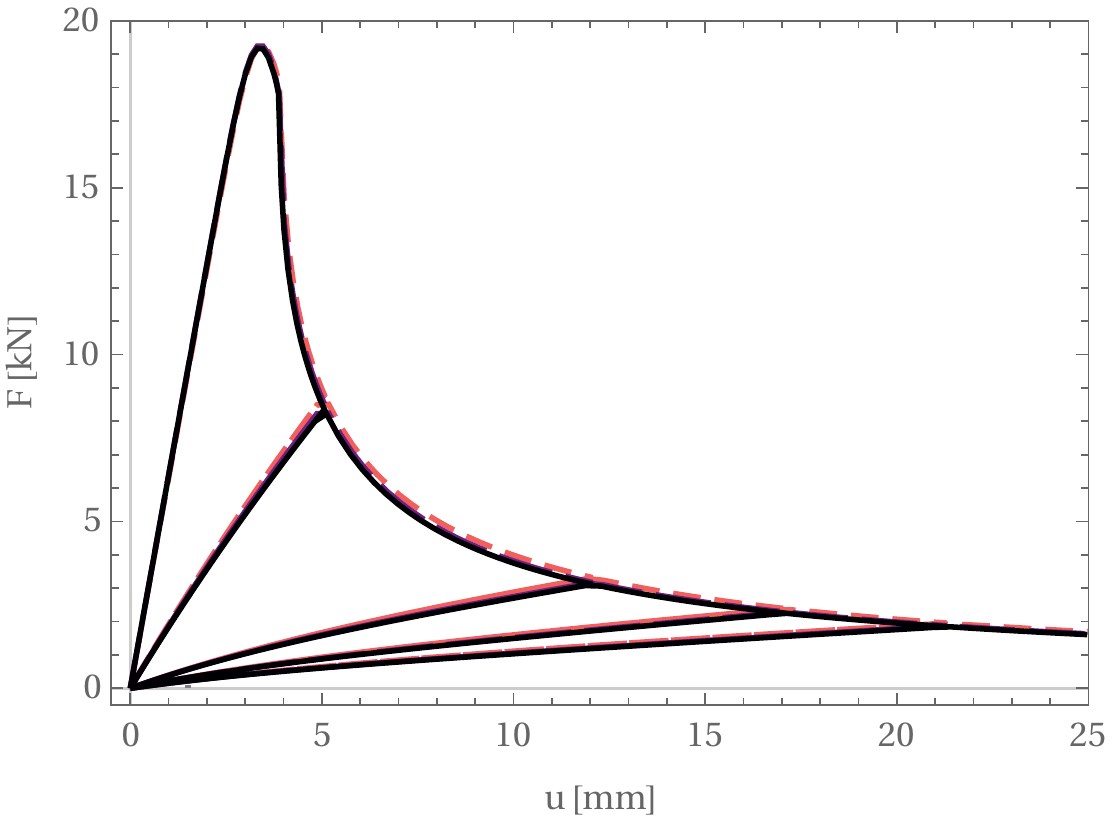}
 \includegraphics[width=0.48\textwidth]{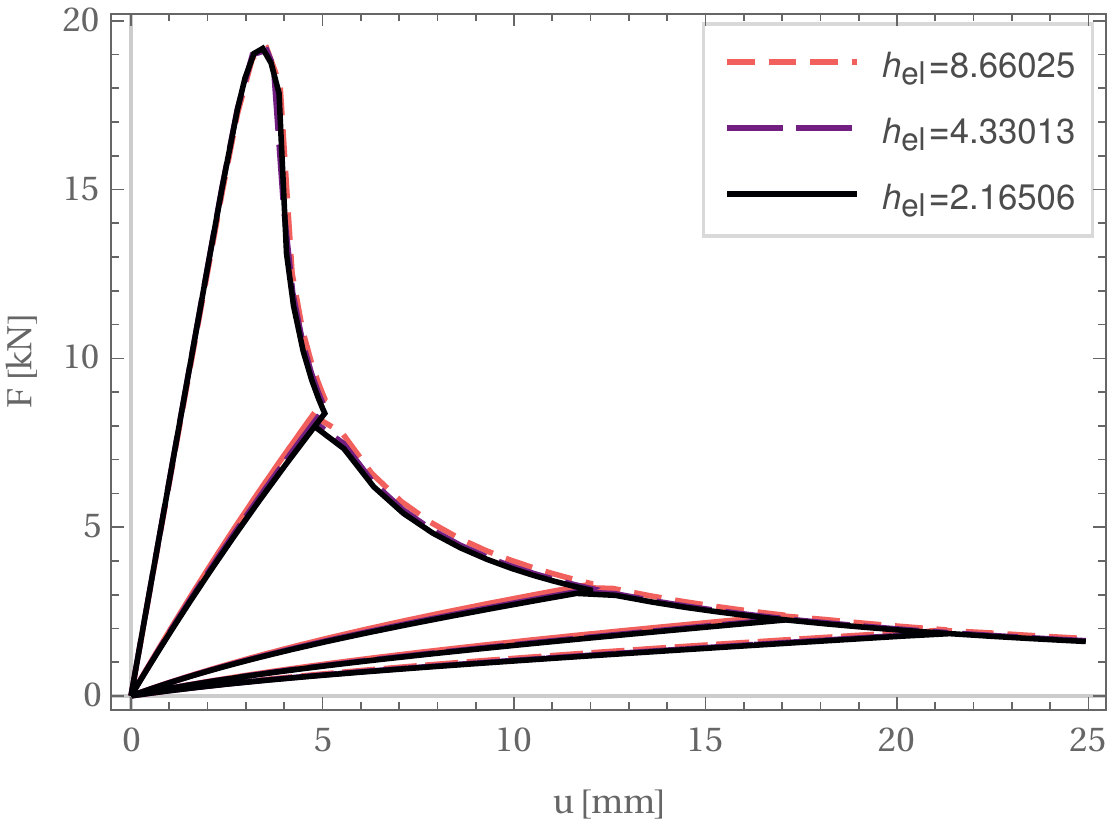}
 \put(-15.5,0){(a)}
\put(-7.5,0){(b)}
\put(-13.3,4.5){\scalebox{0.7}{
\put(0,0.8){$d_0 = 1$ MPa}
\put(0,0.4){$d_1 = 0$ MPa}
\put(0,0){$c_{\phantom{1}} = 100$ Nmm}
\put(0,-0.4){$n_{\mrm{s}} = 500$ }
}}
\put(-5,4.5){\scalebox{0.7}{
\put(0,0.8){$d_0 = 1$ MPa}
\put(0,0.4){$d_1 = 0$ MPa}
\put(0,0){$c_{\phantom{1}} = 100$ Nmm}
\put(0,-0.4){$n_{\mrm{s}} = 200$ }
}}
\put(-11.5,3.25){\includegraphics[width=4cm]{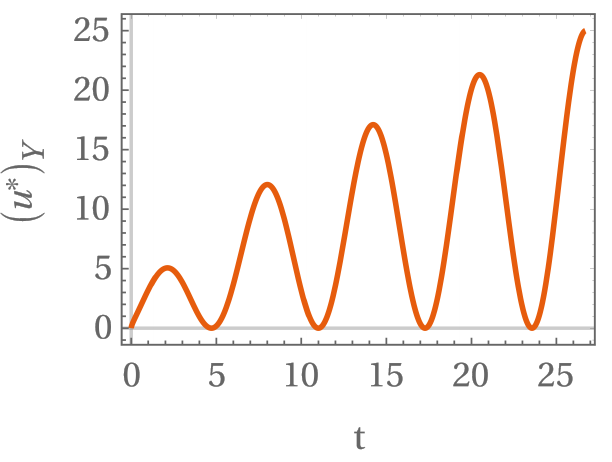}}
\put(-10.5,1.5){\includegraphics[width=2.5cm]{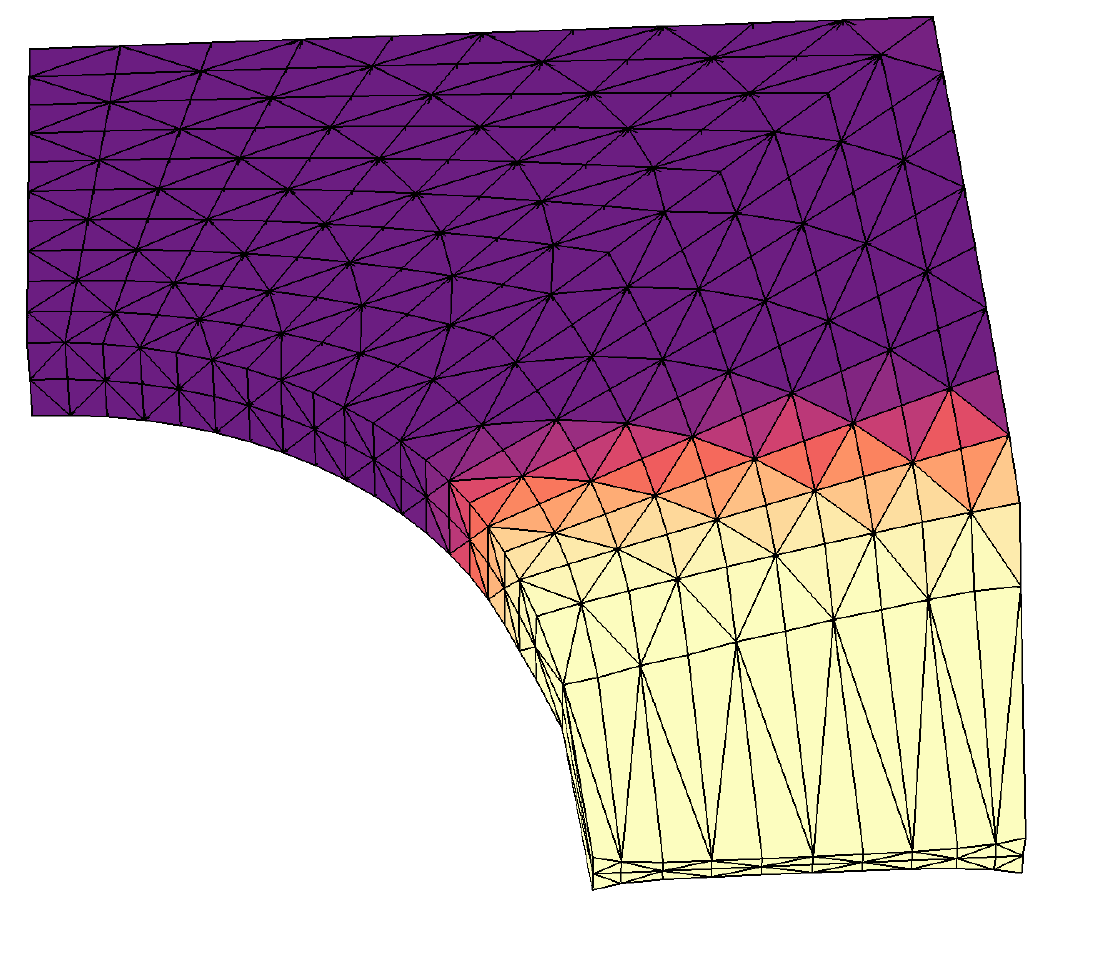}}
\put(-2.65,1.5){\includegraphics[width=2.5cm]{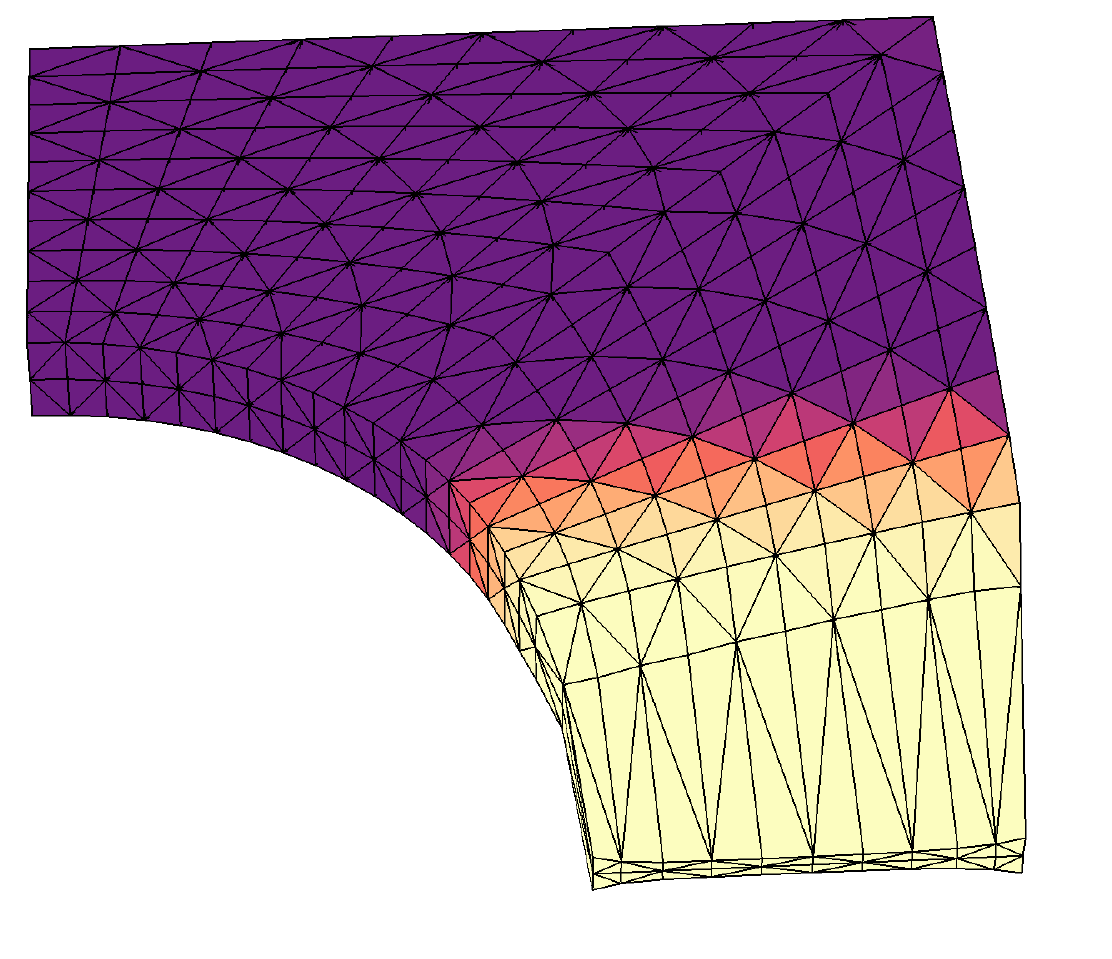}}
}
\end{picture}
\caption{
Force displacement plots for the cyclic prescribed displacement function \eqref{e:cyclic_load_function} for varying number of time steps and damage contourplots corresponding to the final load stage.
((a) $n_{\mrm{steps}}=500$, (b) $n_{\mrm{steps}}=200$).
As can be seen, the resulting mesh-independency turns out to be insensitive to the choice of the time step size. 
}
\label{f:pwh_cyclic_loading}
\end{figure}
In order to analyze the numerical behavior of the proposed formulation and the update algorithm \ref{a:solution_strategy} for loading-unloading scenarios, the prescribed displacement is set to the cyclic function
\eb
u^{\star}_Y(t)=  \frac{t^{0.6} \sin(t) + t^{0.6}}{2 (8.5 \pi)^{0.6}} \ u_Y^{\star,\mrm{max}} \quad\text{with } u_Y^{\star,\mrm{max}}= 25\ \mrm{mm}.
\label{e:cyclic_load_function}
\ee
Corresponding force displacement curves are shown in figure \ref{f:pwh_cyclic_loading} (a) for $n_{\mrm{step}}=500$ load steps and (b) for $n_{\mrm{step}}=200$ load steps, respectively.
In figure \ref{f:pwh_cyclic_loading} (a) the prescribed cyclic displacement function \eqref{e:cyclic_load_function} is also visualized.
By comparing the results depicted in figure \ref{f:pwh_cyclic_loading} (a) to the results of computations with the coarser time discretization of figure \ref{f:pwh_cyclic_loading} (b) it can be observed, that for the latter a similar quality of mesh convergence is present.
Furthermore, despite the larger load steps of the $200$-step computations and resulting therefrom relatively large damage propagation within one Newton loop, no loss of convergence of the iterative solution procedure was observed for the present computations. 
 \section{Conclusion\label{sec:conclu}}
A new mixed finite element formulation for finite strain gradient damage was introduced, where through a Lagrange multiplier constraint compatiblilty between the nodal damage variable and a local history variable was ensured.
The key feature of the constraint term was the incorporation of the damage evolution inequality conditions:
By utilizing the value of the Lagrange multiplier to identify the damage loading and de-loading case and a corresponding update of a local history variable the no-healing inequality was incorporated.
The proposed approach has the advantage that no additional computational resources for the numerical evaluation of the evolution/no-evolution condition and the storage of the values of the history parameter are necessary, since corresponding quantities are already given from the global iterative solution procedure.
Through volume-bubble-enhanced interpolation of the damage variable a formulation yielding rank sufficient tangent matrices for piecewise constant Lagrange mutlipliers was obtained, while removing the necessity for a penalty term.
Further, the discretization enabled static condensation yielding a positive symmetric global matrix with a minimized number of equations being quite similar to competitive penalty approaches.
Numerical tests showed convergence of the solution procedure for severe damage ($D\approx 1$) and cyclic loading conditions.
Mesh independence and improved convergence behavior compared to penalty approaches was shown.
 \makeacknowledgement
 \makebib{references}
 
\begin{appendix}
\section{Appendix \label{app:A}} 

\subsection{Derivation of the Strong Form of the Gradient Damage Problem \label{app:strongf_initial_problem}}
We denote the scalar product by $(\bullet) \scp (\bullet)$, where $(\bullet)$ may be a vector or any $n$-order tensor.
With the definition $\bP\defeq \p{\bF}{\sed}$ from section \ref{ss:cont_damage_mech} the energy balance derived from the potential \eqref{e:total_dissipation_potential} reads:
\eb
\begin{split}
\dot{\varPi} &= \int_{\B} \Big[ \bP \scp \dot{\bF}- \bbf \scp \dot{\bu} \Big] \dX 
- \int_{\dBN} \bt \scp \dot{\bu} \dA \\
&+ \int_{\B} \Big[ \p{\iv}{\sed} \, \dot{\iv} + \nlpar \, \Grad\iv \scp  \Grad\dot{\iv} + \p{\iv}{\diss}\,\dot{\iv} \Big] \dX = 0
\label{e:energy_balance_gradient_damage}
\end{split}
\ee
Here, with $\dot{\bF} = \p{t}{(\Grad\bu+\bone)}=\Grad\dot{\bu}$ and the divergence theorem, the following relations hold:
\begin{align}
 \int_{\B} \bP \scp \Grad\dot{\bu} \dX               &= \int_{\dB} \bP\smpc\nv \scp \dot{\bu} \dA - \int_{\B} \Div{\bP} \scp \dot{\bu} \dX 
 \label{e:div_theorem_u}\\
 \int_{\B} \nlpar \,\Grad\iv \scp \Grad\dot{\iv} \dX &= \int_{\dB} \nlpar\, \Grad\iv\smpc\nv\, \dot{\iv} \dA - \int_{\B} \nlpar \,\Delta\iv \scp \dot{\iv} \dX
 \label{e:div_theorem_iv},
\end{align}
Inserting \eqref{e:div_theorem_u} and \eqref{e:div_theorem_iv} into \eqref{e:energy_balance_gradient_damage} and reordering yields:
\eb
\begin{split}
 \dot{\varPi} &= \int_{\B} \Big[-\Div \bP - \bbf \Big] \scp \dot{\bu}  \dX 
+ \int_{\dBN} \Big[ \bP \smpc \nv - \bt \Big] \scp \dot{\bu} \dA + \int_{\dBD} \bP \smpc \nv \scp \dot{\bu} \dA\\
&+ \int_{\B} \Big[ \underbrace{\p{\iv}{\sed} - \nlpar\, \Delta\iv + \p{\iv}{\diss}}_{\defeq - \yieldf} \Big] \, \dot{\iv} \dX 
+ \int_{\dB} \nlpar\,\Grad\iv \smpc \nv \, \dot{\iv} \dA = 0
\label{e:energy_balance_gradient_damage_strong_form}
\end{split}
\ee
Since \eqref{e:energy_balance_gradient_damage_strong_form} must be fulfilled independent of $\dot{\bu}$ and $\dot{\iv}$ we obtain equations~\eqref{e:equation_set}.

\subsection{Variational Equation Corresponding to the Proposed Lagrangian \label{app:variation_proposed}}

In the continuous setting stationarity of \eqref{e:lagrangian_proposed} is obtained through variation with respect to the solution variables.
Application of the divergence theorem analogously to \eqref{e:div_theorem_u} and \eqref{e:div_theorem_iv} yields the equation
\eb
\begin{split}
\delta L &= \int_{\B} \Big[-\Div \bP - \bbf \Big] \scp \delta{\bu}  \dX 
+ \int_{\dBN} \Big[ \bP \smpc \nv - \bt \Big] \scp \delta{\bu} \dA \\
&+ \int_{\B} \Big[ \underbrace{\p{\iv}{\sed} - \nlpar\, \Delta\iv + \p{\iv}{\diss}}_{\defeq - \yieldf} + \lag \Big] \, \delta{\iv} \dX 
+ \int_{\dB} \nlpar\,\Grad\iv \smpc \nv \, \delta{\iv} \dA \\
&+ \int_{\dB} (\iv- \bar{\iv})\,  \delta\lag \dX=0
\label{e:variational_equation_proposed_strong_form}
\end{split}
\ee
with the test functions belonging to the spaces $(\delta\bu,\delta\iv,\delta\lag)\in \U\times \mathcal{A}\times\mathcal{L}$ and with $\delta\bu\vert_{\dBD}=0$.
From \eqref{e:variational_equation_proposed_strong_form} equations \eqref{e:bal_lin_mom_prop}-\eqref{e:strong_form_constraint} can easily be identified.
\end{appendix}


\begin{thebibliography}{10}

\bibitem{Lem:1984:htu}
J.~Lemaitre.
\newblock How to use damge mechanics.
\newblock {\em Nuclear Engineering and Design}, 1984.

\bibitem{LemCha:1990:mos}
J.~Lemaitre and J.L. Chaboche.
\newblock {\em Mechanics of solid materials}.
\newblock Cambridge University Press, 1990.

\bibitem{Kac:1986:itc}
L.M. Kachanov.
\newblock {\em Introduction to continuum damage mechanics}.
\newblock Mechanics of elastic stability. Springer-Science+Business Media,
  1986.

\bibitem{Mie:dac:1995}
C.~Miehe.
\newblock Discontinuous and continuous damage evolution in ogden-type
  large-strain elastic materials.
\newblock {\em Eur. J. Mech. A-Solid}, 14:697–720, 1995.

\bibitem{MenSte:2001:ata}
A.~Menzel and P.~Steinmann.
\newblock A theoretical and computational framework for anisotropic continuum
  damage mechanics at large strains.
\newblock {\em Int. J. Solids Structures}, 38(52):9505–9523, 2001.

\bibitem{BalSchGro:sod:2006}
D.~Balzani, J.~Schröder, and D.~Gross.
\newblock Simulation of discontinuous damage incorporating residual stresses in
  circumferentially overstreched atherosclerotic arteries.
\newblock {\em Acta Biomaterialia}, 2006.

\bibitem{BalBriHol:2012:cff}
D~Balzani., S.~Brinkhues, and G.~Holzapfel.
\newblock Constitutive framework for the modeling of damage in collagenous soft
  tissues with application to arterial walls.
\newblock {\em Comp. Methods Appl. Mech. Engrg.}, 2012.

\bibitem{GueMie:2011:oed}
E.~Gürses and C.~Miehe.
\newblock On evolving deformation microstructures in non-convex partially
  damage solids.
\newblock {\em Journal of Mechanics and Physics of Solids}, 59:1268–1290,
  2011.

\bibitem{BalOrt:2012:riv}
D.~Balzani and M.~Ortiz.
\newblock Relaxed incremental variational formulation for damage at large
  strains with application to fiber-reinforced materials and materials with
  truss-like microstructures.
\newblock {\em Int. J. Numer. Meth. Eng.}, 92:551–570, 2012.
\newblock .DOI: 10.1002/nme.4351.

\bibitem{SchBal:2016:riv}
T.~Schmidt and D.~Balzani.
\newblock Relaxed incremental variational approach for the modeling of
  damage-induced stress hysteresis in arterial walls.
\newblock {\em J. of the Mech. Behavior of Biomedical Materials}, 58:149–162,
  2016.

\bibitem{Mie:2011:amf}
C.~Miehe.
\newblock A multi-field incremental variational framework for gradient-extended
  standard dissipative solids.
\newblock {\em Journal of Mechanics and Physics of Solids}, 59:898–923, 2011.

\bibitem{SchJunHac:2020:vro}
S.~Schwarz, P.~Junker, and K.~Hackl.
\newblock Variational regularization of damage models based on the emulated
  {RVE}.
\newblock {\em Cont. Mech. and Thermodyn.}, 2020.

\bibitem{KoeNeuMelPetBal:2022:aco}
M.~Köhler, T.~Neumeier, J.~Melchior, D.~Petersheim, and D.~Balzani.
\newblock Adaptive convexification of microsphere-based incremental damage for
  stress and strain softening at finite strains.
\newblock {\em Acta Mechanica}, (in press).

\bibitem{PeeBorBreVre:1996:ged}
R.H.J. Peerlings, A.M. de~Borst, R. an~Breckelmans, and J.H.P. de~Vree.
\newblock Gradient enhanced damage for quasi-brittle materials.
\newblock {\em International Journal for Numerical Methods in Engingeering},
  39:3391–3403, 1996.

\bibitem{PlaBarMisTim:2021:mbe}
L~Placidi., E.~Barchiesi, A.~Misra, and D.~Timofeev.
\newblock Micromechanics-based elasto-plastic-damage energy formulation for
  strain gradient solids with granular microstructure.
\newblock {\em Continuum Mech. Thermodyn.}, 33:2213--2241, 2021.

\bibitem{JunSchJanHac:2019:afa}
P.~Junker, S.~Schwarz, D.R Jantos, and K.~Hackl.
\newblock A fast and robust numerical treatment of a gradient-enhanced model
  for brittle damage.
\newblock {\em Int. J. Multisc. Comp. Eng.}, 17(2):151–180, 2019.

\bibitem{VogJun:2019:aah}
A.~Vogel and P.~Junker.
\newblock Adaptive and highly accurate numerical treatment for a
  gradient-enhanced brittle damage model.
\newblock {\em Int. J. Numer. Meth. Eng.}, 2019.

\bibitem{JunRieBal:2021:ent}
P.~Junker, J.~Riesselmann, and D.~Balzani.
\newblock Efficient and robust numerical treatment of a gradient-enhanced
  damage model at large deformations.
\newblock {\em International Journal of Numerical Methods in Engineering},
  123:774--793, 2022.

\bibitem{LieSteBen:2001:tac}
T.~Liebe, P.~Steinmann, and A.~Benallal.
\newblock Theoretical and computational aspects of a thermodynamically
  consistent framework for geometrically linear gradient damage.
\newblock {\em Comput. Methods Appl. Mech. Engrg.}, 2001.

\bibitem{DimHac:2008:amf}
B.J. Dimitrijevic and K.~Hackl.
\newblock A method for gradient enhancement of continuum damage models.
\newblock {\em Techn. Mech.}, 28(1), 2008.

\bibitem{WafPolMenBla:2013:age}
T.~Waffenschmidt, C.~Polindara, A.~Menzel, and S.~Blanco.
\newblock A gradient-enhanced large-deformation continuum damage model for
  fibre-reinforced materials.
\newblock {\em Comp. Methods Appl. Mech. Engrg.}, 268:801–842, 2013.

\bibitem{For:2009:maf}
S.~Forest.
\newblock Micromorphic approach for gradient elasticity, viscoplasticity and
  damage.
\newblock {\em Journal of Engineering Mechanics}, 135(3), 2009.

\bibitem{MieWelHof:2010:tcp}
C.~Miehe, F.~Welschinger, and M.~Hofacker.
\newblock Thermodynamically consistent phase-field models of fracture:
  Variational principles and multi-field fe implementations.
\newblock {\em Int. J. Numer. Meth. Eng.}, 83:1273–1311, 2010.

\bibitem{MieHofWel:2010:apf}
C.~Miehe, M.~Hofacker, and F.~Welschinger.
\newblock A phase field model for rate-independent crack propagation: Robust
  algorithmic implementation based on operator splits.
\newblock {\em Comp. Methods Appl. Mech. Eng.}, 199:2765–2778, 2010.

\bibitem{GerDeL:2019:opi}
T.~Gerasimov and L.~{De Lorenzis}.
\newblock On penalization in variational phase-field models of brittle
  fracture.
\newblock {\em Comput. Methods Appl. Mech. Engrg.}, 354:990–1026, 2019.

\bibitem{Wri:2008:nfe}
P.~Wriggers.
\newblock {\em Nonlinear Finite Element Methods}.
\newblock Springer, 2008.

\bibitem{KuhTuc:1951:np}
H.W. Kuhn and A.W. Tucker.
\newblock Nonlinear programming.
\newblock In Jerzy Neyman, editor, {\em Proceedings of the Berkeley Symposium
  on Mathematical Statistics and Probability}, volume~2, page 481–492.
  University of California Press, 1951.

\bibitem{BofBreFor:2013:mfe}
D.~Boffi, F.~Brezzi, and M.~Fortin.
\newblock {\em Mixed Finite Element Methods and Applications}.
\newblock Springer, 2013.

\bibitem{Bra:2007:fel}
D.~Braess.
\newblock {\em Finite Elemente}.
\newblock Springer, 2007.
\newblock pp.162-163.

\bibitem{ArnBre:1985:man}
D.~Arnold and F.~Brezzi.
\newblock Mixed and nonconforming finite element methods: implementation,
  postprocessing and error estimates.
\newblock {\em Math. Anal. Numer.}, 19(1):7–32, 1985.

\bibitem{RieBal:2022:fef}
J.~Riesselmann and D.~Balzani.
\newblock Finite element formulations for gradient damage at finite strains.
\newblock In {\em Current Trends and Open Problems in Computational Mechanics},
  page 443–452. Springer, 2022.

\end{thebibliography}
\end{document}